\documentclass{aa}
\usepackage{graphicx}
\usepackage{txfonts}
\usepackage{natbib}
\usepackage{rotating}
\usepackage{color}
\usepackage{longtable,lscape}
\begin{document}

\title{Where does the gas fueling star formation in BCGs originate?}

\subtitle{}


\author{S. Molendi\inst{1}, P. Tozzi\inst{2}, M. Gaspari\inst{3}, S. De Grandi\inst{4},
F.Gastaldello\inst{1,5}, S.Ghizzardi\inst{1} and M.Rossetti\inst{6,1} }

\offprints{S. Molendi \email{silvano@iasf-milano.inaf.it}}

\institute{INAF - IASF Milano, via E. Bassini 15 I-20133 Milano, Italy \and
INAF - Osservatorio Astrofisico di Firenze, Largo Enrico Fermi 5, I-50125, Firenze, Italy \and
Department of Astrophysical Sciences, Princeton University, Princeton, NJ 08544, USA; Einstein and Spitzer Fellow \and
INAF - Osservatorio Astronomico di Brera, via E.Bianchi 46, 23807 Merate, Italy \and
Department of Physics and Astronomy, University of California at Irvine, 4129 Frederick Reines Hall, Irvine, CA 92697-4575, USA \and
Dipartimento di Fisica dell'Universit\`a degli Studi di Milano, via Celoria 16, I-20133, Milan, Italy
}

\date{today}


\abstract
{}
{We investigate the relationship between X-ray cooling and star formation in
brightest cluster galaxies (BCGs).}
{We present an X-ray spectral analysis of the inner regions, 10-40 kpc, of six nearby cool core
clusters ($z<0.35$) observed with  {{\sl Chandra}} ACIS.  This sample is selected on the basis of
the high star formation rate (SFR) observed in the BCGs.
We restrict our search for cooling gas to regions that are roughly cospatial with the starburst.
We fit single- and multi-temperature {{\tt mkcflow}} models to  constrain  the
amount of isobarically cooling intracluster medium (ICM).}
{
We find that in all clusters, below a threshold temperature ranging between 0.9 and 3 keV,
only upper limits can be obtained.
In four out of six objects, the upper limits are significantly below the SFR and in two, namely A1835 and A1068, they are less than a tenth of the SFR.
}
{{
Our results suggests that a number of mechanisms  conspire to hide
the cooling signature in our spectra. In a few  systems the lack of
a cooling signature may be attributed to a relatively long delay time between the
X-ray cooling and the star burst. However, for A1835 and A1068, where the X-ray
cooling time is shorter than the timescale of the starburst, a possible explanation
is that the region where gas cools out of the X-ray phase extends to very large radii,
likely beyond the core of these systems.
 }}


\titlerunning{Where does the gas fueling star formation in BCGs come from?}

\authorrunning{Molendi et al.}

\maketitle

\section{Introduction}

Most of the baryonic matter in galaxy clusters resides in
the form of virialized hot gas \citep{2003Lin,2013Gonzalez} that emits in the X-ray band
via thermal bremsstrahlung.   The short cooling
time associated with the  high-density regions in cluster cores, immediately led to the
conclusion that a massive cooling flow was developing in the ICM of most
clusters \citep{1976Silk,1977Cowie,1977Fabian,1978Mathews}.
On the basis of the isobaric cooling flow model \citep{1977Fabian,1994Fabian}, it was estimated that typical
cooling flows could develop mass cooling rates in the range $ 100 - 1000 M_{\sun}$ yr$^{-1}$.
However, the lack of massive star formation events and of large reservoirs of cold gas
at the center of galaxy clusters
has cast  some doubts on the hypothesis of a complete cooling of the ICM in cluster cores.
Indeed, if a significant fraction of the ICM  is deposited, it should cool
through all the phases and, in a quasi-stationary scenario, a noticeable amount of cooling gas
should be observed at all temperatures in the X-ray range, from the ambient temperature down to
the lowest detectable temperature (about $0.15$ keV).

The perspective changed radically  fifteen years ago when
XMM-Newton observations, carried out  with EPIC and with RGS, revealed a lack of cool gas at
temperatures lower than about one-third of the cluster virial temperature
 \citep{2001Molendi,2001Tamura,2003Peterson}.  As a consequence, the mass cooling rates have been
revised towards values more than one order of magnitude lower \citep[see][]{2006Peterson}.
These observations necessarily imply the existence of some process that heats
the gas and suppresses its cooling.  Among the many mechanisms investigated so far,
AGN feedback is considered  the most plausible \citep[see][]{2012Fabian}.  Although the
detailed underlying physics is not fully understood, AGN outbursts can inject sufficient
amounts of energy into the ICM to
offset cooling \citep[e.g.,][]{2005McNamara,2015Gaspari}.
The relativistic electrons in jets associated with the
central cluster galaxy are able to carve large cavities into the ICM.  The free energy associated
with these ``bubbles,, is plausibly transferred into the ICM and thermalized through turbulence
\citep[see][]{2012McNamara}.
This feedback mechanism has been observed in its full complexity in a few nearby clusters like
Perseus \citep{2003aFabian,2006Fabian,2011Fabian}, Hydra A \citep{2000McNamara},
and a few others at large distances \citep[see][]{2011Blanton}.

In addition, some of the large elliptical galaxies often found at the minimum of the cluster
potential well (BCG) contain significant amounts of cold gas \citep[among the most recent
results, see][]{2014McNamara} and exhibit signs of ongoing star formation at a level
of $10-100 M_{\sun}$ yr$^{-1}$ \citep{2012Hoffer}.
The star formation rate (SFR) of BCGs has been found to
be generally consistent with the mass cooling rate $\dot M$ measured with X-ray spectroscopy in cool cores
\citep{2006Rafferty}; we note, however, that \citet{2008Odea} have found that cooling rates are in excess
of star formation rates by a factor of a few.  A correlation between cool cluster cores and IR measured
SFR in BCGs has  recently been confirmed  \citep{2012Rawle}. These results have generally been interpreted
as evidence that  star formation is driven and fueled by the cooling of the ICM.



In this paper we revisit this relation by improving the measurement of the mass
cooling rates in cool cores.  We start from the fact that
the emission measure of the coldest gas ($<1 $ keV) in cooling flow models is a few percent
of the total X-ray emission in the core region, so that the measurement of mass cooling rate
based on X-ray spectroscopy is dominated by the emission of
gas at relatively high temperatures (about 2 keV).  The presence of gas in the range $0.5$-$1$
keV is mostly associated with the Fe XVII emission lines, which is the most
reliable diagnostic for gas in this temperature range.  Although Fe XVII emission
lines have been detected in some cases, the amount of cold gas associated with that emission
is still quite uncertain owing to the strong continuum produced by much hotter gas which
unavoidably overwhelms the line emission.

In this paper we  address a specific issue: Is the amount of gas cooling
in the region where star formation is occurring sufficient to fuel this process?
Within this framework,  angular resolution becomes a critical factor in order to ascertain the presence of the
coldest components of a cooling flow; in several cases it is more relevant than spectral resolution,
and CCD data can be as effective as grating data despite its modest
spectral resolution.

In this work we exploit the angular resolution of {\sl Chandra} ACIS-S and ACIS-I to avoid
as much as possible the hot gas surrounding  the core region.
We consider six  nearby ($z<0.35$) cool core clusters observed
with {\sl Chandra}, whose BCG exhibit the highest SFR.  For all these clusters, the mass cooling rates
have been previously estimated from X-ray spectral analysis, but
in all cases within regions much larger than those considered here (10-40 kpc
centered on the BCG).  Our goal is to robustly  estimate the mass cooling rate
$\dot M$ or, alternatively,  to put a robust upper limit to $\dot M$, and finally to compare this to the
observed SFR in the same region.

The paper is organized as follows. In \S 2 we describe the sample selection.
In \S 3 we describe the data reduction and analysis, and in
\S 4 we provide the measurements of the mass cooling rate in each cluster of our sample.
In \S 5 we go through a detailed evaluation of possible systematic effects that can
potentially hide the presence of cooling gas or interfere with its  measurement.
The final result is a robust evaluation of mass cooling rates $\dot M$
which includes systematic effects.
In \S 6 we discuss our results, and  finally, in \S 7 we summarize our conclusions.
Throughout the paper, we adopt a $\Lambda$CDM cosmology
with $\Omega_0=0.3, \Omega_\Lambda=0.7$, and $H_0=70$.
Quoted errors always correspond to a 1 $\sigma$ confidence level.

\section{Sample selection}

\begin{table*}
\centering
\caption{\label{tab:zobs}Galaxy clusters selected in our sample, ranked from the highest to the
lowest measured SFR in the BCG.}
\begin{tabular}{|c|c|c|c|c|c|c|}
\hline
Name   & $z$ & SFR & SFR Orig.$\mathrm{^a}$ & SFR Ref.$\mathrm{^b}$ &$M_{\rm mol}\,^\mathrm{^c}$ & $M_{\rm mol}$ Ref.$\mathrm{^d}$ \\
       &                   &$M_\odot$yr$^{-1}$         &    & & $10^{10}M_\odot$ &   \\
\hline
A1835                           & 0.252       &       146       &  IR SED   &  Ra12  & 5.0  & Mn14\\
RXCJ1504.1                      & 0.215       &       140       & H$_\alpha$&  Og10  &  -   &  -   \\
RX J1532.9+3021             & 0.345       &       110       & 70 $\mu$m &  Ho12  & 12.5$^\mathrm{^e}$ & Ed01\\
A1068                           & 0.139       &      99.3       &  IR SED   &  Ra12  & 3.7$^\mathrm{^f}$  & Sa03\\
ZW 3146                 & 0.291       &      93.1       &  IR SED   &  Ra12  &  8.0$^\mathrm{^e}$ & Ed01\\
Z0348                           & 0.254       &      32.6       & IR SED    &  Ra12  & -    &  -   \\
\hline
\end{tabular}
\begin{list}{}{}
\item[Notes:]
   $\mathrm{^a}$~band or method employed to measure SFR;
   $\mathrm{^b}$~reference for SFR: Ra12 = \citet{2012Rawle}; Og10 = \citet{2010Ogrean}; Ho12 = \citet{2012Hoffer};
   $\mathrm{^c}$~molecular gas mass, $M_{\rm mol}$;
   $\mathrm{^d}$~reference for $M_{\rm mol}$: Mn14 = \citet{2014McNamara}; Ed01 = \citet{2001Edge};
   Sa03 = \citet{2003Salome};
   $^\mathrm{^e}$ molecular gas masses originally reported in \citet{2001Edge} have been recomputed using the cosmological parameters described in \S 1 and assuming an average Galactic value for the  CO to molecular hydrogen conversion factor \citep{2013Bolatto}, i.e., $X_{\rm CO}=2\times 10^{20}$cm$^{-2}$(K km s$^{-1}$)$^{-1}$;
   $^\mathrm{^f}$ molecular gas mass originally reported in \citet{2003Salome} has been recomputed assuming an average Galactic value for the  CO to molecular hydrogen conversion factor \citep{2013Bolatto}, i.e., $X_{\rm CO}=2\times 10^{20}$cm$^{-2}$(K km s$^{-1}$)$^{-1}$.

\end{list}
\end{table*}

Our goal is to investigate the relation between the
mass cooling rate and star formation rate in cool core (CC) clusters.
We  selected those CC clusters with high-quality X-ray data and the
highest measured SFR.  This choice allows us to investigate the relation between BCG SFR and $\dot M$
starting from the clusters where the expected $\dot M$ should be highest\footnote{see \S\ref{sec:dis} for
a discussion of possible delays between X-ray cooling and star formation events.}.

We search in the Chandra archive as of June 2014 for clusters with ACIS-I or ACIS-S
observations longer than 9 ks. We then select those clusters that have published
SFR in excess of 30 $M_\odot$yr$^{-1}$.  When there is  more than one measurement of the SFR,
we prefer measurements that are based on infrared data, whenever available, since UV and
optical measures can significantly underestimate obscured star formation \citep[see][]{2012Rawle}.
We also favor estimates based on infrared spectral energy distribution
(IR SED) fitting, rather than on single-band measurements, as they can properly account for the
contribution from the central AGN which, in many cases, can be relevant.
For systems where this method has been used,
the uncertainties on the star formation rates are expected to be modest, below $\sim 20\%$
 \citep[see][]{2012Rawle}.
We finally apply a cut in redshift,  choosing systems with $ z < 0.35$.
The upper limit is chosen in order to avoid clusters where a large part of the cold gas emission
is redshifted out of the {\sl Chandra} energy range used for this analysis ($0.5-8$ keV).
We note  that one of the strongest candidates that harbors a massive cooling flow,
SPT-CL J2344-4243 \citep[the Phoenix cluster, see][]{2012McDonald},
is excluded from this sample owing to its high redshift $z\sim 0.6$.  A dedicated paper,
based on a deep XMM-Newton observation, has recently been published by our group
\citep{2015bTozzi}.

Our final sample is rather small, and consists of six systems, most of which are well known.
In Table~\ref{tab:zobs} we list our objects and include  redshift, measured SFR in the BCG,
the method used to measure the SFR, and measured cold molecular gas, when available.  This sample
is not meant to be complete or representative, but simply satisfies the requirements needed to
perform our analysis: {\sl Chandra} ACIS data, presence of a cool core, and a reliable measurement of
SFR in the BCG larger than 30 $M_\odot$yr$^{-1}$.

\begin{table*}
\centering
\caption{\label{tab:obsid} Summary of {\sl Chandra} observations used in this work}
\label{tab:data}
\begin{tabular}{|c|c|c|c|c|}
\hline
Name   & Obsid   & Instrument &  Effective exptime$^a$ & Observing Mode  \\
           &             &             &           s                  &                            \\
\hline
                     &           6880      & ACIS-I &  117700   & VFAINT\\
A1835           &     6881          & ACIS-I & 36200    & VFAINT \\
                     &      7379            & ACIS-I & 39300    & VFAINT\\
\hline
RXCJ1504     &                  5793      &  ACIS-I      &   39000      & VFAINT \\
                    &           4935      &  ACIS-I      &   13300      &  FAINT \\
\hline
RX J1532             &  14009           & ACIS-S         &    88000     &  VFAINT\\
\hline
A1068            &  1652                & ACIS-S                 &    26600     &   FAINT    \\
\hline
 ZW3146        &        909     &  ACIS-I        &    45600    &  FAINT  \\
                          &     9371    &  ACIS-I        &    39800    &  VFAINT  \\
\hline
Z0348          &        10465   & ACIS-S         & 47800        &   VFAINT  \\
\hline
\end{tabular}
\begin{list}{}{}
\item[Notes:]
   $\mathrm{^a}$~effective exposure time after data reduction.
   \end{list}
\end{table*}

\section{Data reduction and analysis}

\subsection{Data reduction}

We performed a standard data reduction starting from the level 1 event
files, using the {\tt CIAO 4.6} software package, with
version ({\tt CALDB 4.6.3}) of the {\sl Chandra} Calibration Database.
All the observation ID numbers (Obsid) and with the corresponding instruments, effective exposure times after data reduction,
and the data acquisition modes are listed in Table \ref{tab:data}.  When
observations are taken in the VFAINT mode, we run the task {\tt
acis$\_$process$\_$events} to flag background events that are most
likely associated with cosmic rays and remove them.  With this procedure, the ACIS particle background can be
significantly reduced compared to the standard grade selection.  The
data is filtered to include only the standard event grades 0, 2, 3, 4,
and 6.  We checked visually for hot columns left after the standard
reduction.  For exposures taken in VFAINT mode, there are practically
no hot columns or flickering pixels left after filtering out bad events.
We also apply CTI correction to the ACIS-I data.  We finally filter time intervals with high background by performing a
$3\sigma$ clipping of the background level using the script {\tt
analyze\_ltcrv}.  The final effective exposure times are generally very close to the
original observing time.  We note that our data reduction is not
affected by possible undetected flare since we are able to compute
the background in the same observation from a large source-free region around the cluster, thus
taking into account any possible spectral distortion of the background
itself induced by the flares.  However, given the small source extraction regions and the high
surface brightness of the core, the background correction is always negligible.

We select regions from which X-ray spectra are extracted in such a way to include the entire star
forming region.  Indeed, choosing regions that do not fully encompass the SF region
may  introduce a bias and lower the cooling rate, while in the opposite case the sensitivity to the cooler X-ray emitting gas will
be significantly reduced owing to the presence of larger amounts of ambient gas.  However, in several cases the
star forming region is not precisely identified.
Therefore,
to show that our results are not sensitive to the size of the extraction regions,
we always analyze the X-ray emission within two different regions selected from a suite of four circles
with radii of 10, 20, 30, and 40 kpc.
All the circular regions are centered on the position of the BCG or of the main starburst, when not coincident with the BCG,
as obtained from optical data.  Details on the specific selections are provided in the subsections describing
the analysis of individual systems.  We exclude the inner 2 arcsec when an unresolved X-ray AGN is found in the BCG.
When a merged spectrum from multiple exposures is extracted, the response matrix files and ancillary
response files are computed for each Obsid and then added together weighted by the corresponding exposure
time.  However, three of our clusters are observed in single exposures.  Finally,
the results for ZW3146 and RXCJ1504 are obtained by performing the combined fit of the spectra from
two different Obsid without merging them due to the different observing modes (FAINT and VFAINT).

\subsection{Fitting method}

Our goal is to constrain the mass cooling rate in the projected regions  10-40 kpc in size centered
on the BCG or on the star forming regions.
We use the model {\tt mkcflow} \citep{1988Mushotzki}, which is based on a combination of
single-temperature {\tt mekal} models \citep{1985Mewe,1986Mewe,1992Kaastra,1995Liedahl}
according to the expectation of the isobaric cooling flow model.  This model assumes a unique
mass cooling rate throughout the entire temperature range probed by the data.
The actual situation may be more complex;  some of the gas above a given
temperature threshold may cool at a relatively high rate, consistent with the isobaric
cooling assumption.  Instead, mass cooling associated with colder gas may be much lower.
Indeed, grating spectra of cool cores are traditionally fitted with an isobaric cooling flow
model with a cutoff temperature below which there is no gas \citep{2006Peterson}.
Therefore, to explore more complex scenarios, we also measure the cooling rate in several
temperature intervals by using a set of  {\tt mkcflow} models in contiguous temperature bins.

In practice, we measure the spectral mass deposition rate in
two different ways:

\begin{itemize}

\item a cooling flow model {\tt mkcflow} coupled to one single-temperature  {\tt mekal}
component.  The upper temperature of the {\tt mkcflow} component is frozen to $3$ keV, while the
temperature of the {\tt mekal} component is left free.  The lowest temperature of the {\tt mkcflow} component
it frozen to $0.15$ keV.  By setting the minimum temperature to $0.15$ keV, we can interpret
the normalization of the {\tt mkcflow} model as the maximum cooling rate allowed
for an isobaric cooling flow across the entire temperature range (from $3$ to $0.15$ keV).
By setting the maximum temperature of the {\tt mkcflow} model to 3 keV, we  include in this
component all the phases that contribute significantly to the Fe-Lshell emission.
A minimum temperature of 4 keV is set for the {\tt mekal} component to avoid superposition
with the {\tt mkcflow} component.
The metal abundance of the {\tt mkcflow}  and  {\tt mekal} models are tied together.


\item A set of {\tt mkcflow} models whose lower and upper temperatures are tied
together in order to avoid overlap.  Here the normalization of each component provides
the mass cooling rate of the gas in the corresponding temperature interval.
The upper and lower temperatures are frozen to the following values:
$0.15-0.25$, $0.25-0.45$, $0.45-0.9$, $0.9-1.8$, and $1.8-3.0$.
As for the single {\tt mkcflow} model, by setting the maximum temperature to 3 keV we  include in the
 {\tt mkcflow} components all the phases that contribute significantly to the Fe-Lshell emission.
A  {\tt mekal} component accounts for the hotter gas; a minimum temperature of 4 keV is set to avoid superposition
with the {\tt mkcflow} components.
 The metal abundance values of the various {\tt mkcflow} components
are tied to one another, while the metal abundance of the {\tt mekal} component is left free.

\end{itemize}

It is worth noting that
above $3$ keV, a single {\tt mekal} model can account for the presence of several hot components
since it is not possible to identify them in the spectral analysis \citep[see][]{2004Mazzotta}.
Another possible approach would be to link the highest temperature of the
last {\tt mkcflow} model to that of the {\tt mekal} model.  However, whenever
we repeat our fits linking the highest {\tt mkcflow} temperature to the {\tt mekal} component,
we find only negligible changes in the best-fit values of the mass cooling rate.

The Galactic absorption, $N_H$, is frozen to the value obtained from the radio map of
\citet{2005LAB} at the position of the cluster.  We will discuss in \S 5 the effect
of leaving the $N_H$ value free to vary under reasonable assumptions.
Finally, the redshift is frozen to the best-fit value obtained by fitting the K$\alpha$ line
emission complex from H-like and He-like Fe after checking that this value is in
agreement with the optical value.

Typically, in the soft energy range (0.5-2 keV), the fraction of the signal
associated with the cold ($kT < 3$ keV) components in our models is below 10\%,
and in several instances below 1\%. Clearly at these very low levels systematic
effects associated with our limited knowledge of the Chandra effective areas
can play an important role. In \S 5 we will present a simple method to
derive estimates of the systematic uncertainties on mass cooling rates.
The use of the {\tt mkcflow} model allows us to directly obtain the best-fit  mass cooling rate
$\dot M$.
Finally, the quoted error bars correspond to a $1 \sigma$ confidence level.

\section{Measurement of mass cooling rate in individual systems}

In this section we  describe the analysis of the cool core regions at different radii
in each cluster.   We  provide full details for A1835, and refer to this analysis for the
remaining clusters.

\subsection{Abell 1835}

\begin{figure*}
\includegraphics[width=8cm]{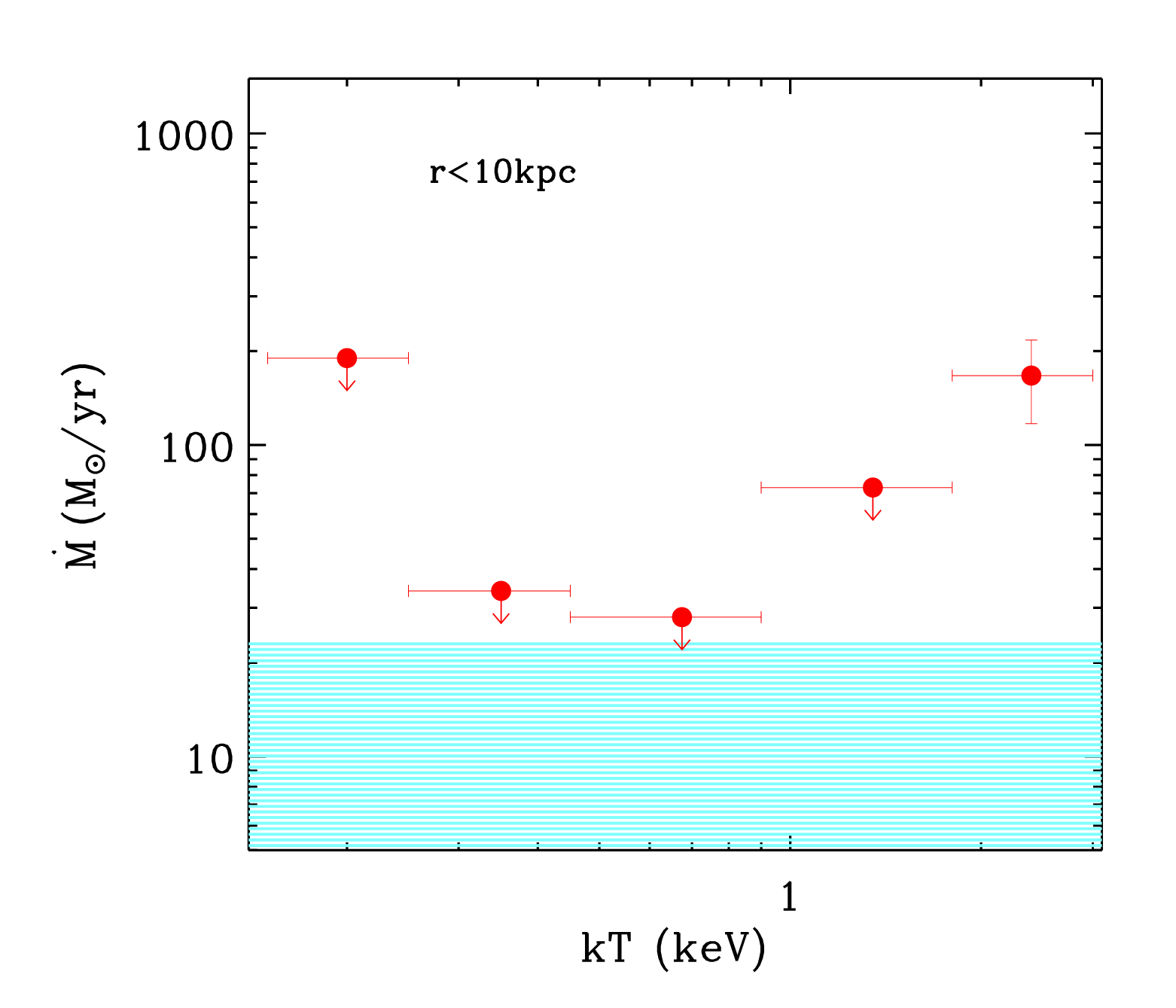}
\includegraphics[width=8cm]{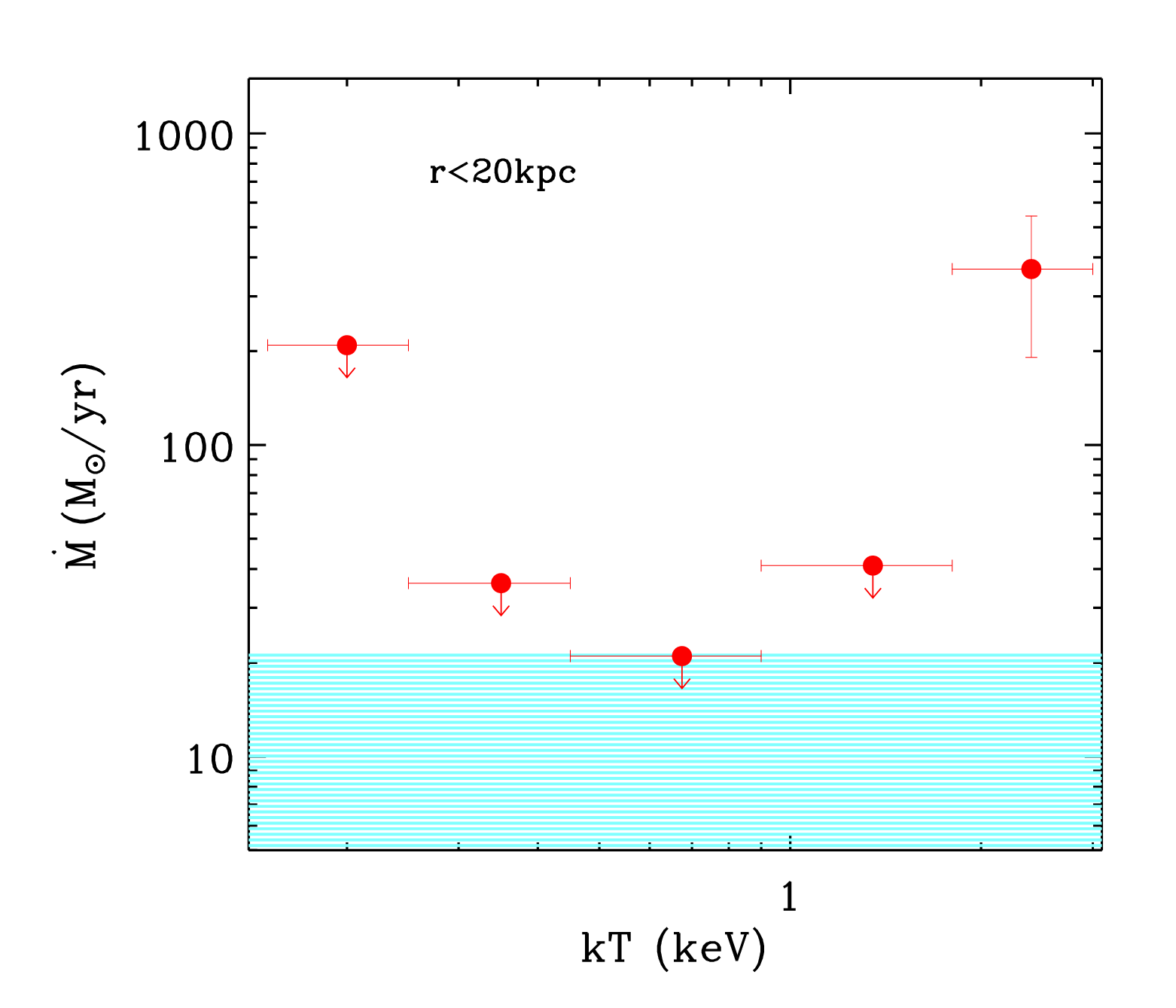}
\caption{\label{A1835mdot}Measured cooling rate as a function of temperature range. {\sl Left}: Red circles show the values of $\dot M$ measured in different temperature bins in the inner 10 kpc of A1835.  Error bars correspond to $1 \sigma$,
arrows indicate 95\% confidence upper limits.  The shaded area is the $ 1 \sigma$ uncertainty or the 95\% upper limit
interval of the global mass cooling rate obtained with a single {\tt mkcflow} model in the temperature
range $0.15-3.0$ keV.
{\sl Right}: Same as left panel for the inner 20 kpc.}
\end{figure*}

A1835 is one of the first clusters in which the lack of cold gas below $\sim$ 1 keV was unambiguously observed
thanks to CCD \citep{2001Molendi} and grating \citep{2001Peterson} spectroscopy with XMM-Newton.  In the latter work,
an upper limit of $200 M_{\sun}$ yr$^{-1}$ at the 90\% confidence level was obtained.
A1835 also harbors a radio AGN, which is responsible for cavities clearly seen in the X-ray surface brightness near
 the center.  For these reasons, A1835 is considered an archetypical cool core cluster regulated by radio-mode
feedback \citep{2014McNamara}.

Abell 1835 was observed with ACIS-S for a total of 30 ks of effective exposure in 1999/2000, and with ACIS-I
for a total of 193 ks in 2005/2006.  The ACIS-S data were taken in the FAINT mode, and date back
to the early epochs of {\sl Chandra} operation.  Here we focus only on the deeper ACIS-I exposures
taken in the VFAINT mode, and which contain most of the information.  The redshift of the cluster
is measured with the X-ray spectral analysis of the global emission, and is found to be
$z_X=0.251_{-0.002}^{+0.001}$.   This value is in good agreement with the
optical value $z_{opt}=0.2523$ \citep{1999Crawford}.  At this redshift  the angular diameter scale corresponds
to $3.91$ kpc/arcsec.  Therefore, 10 kpc corresponds to 5.2 physical pixels on ACIS-I
(one pixel corresponds to $0.492$ arcsec) and
$20$ kpc to $10.4$ pixels.  Given the exquisite resolution of
{\sl Chandra} at the aimpoint, we neglect effects due to the point spread function.
We choose the center to be on the optical position of the BCG, which is
$RA_{BCG}=14$:$01$:$02.0$, $DEC_{BCG}=+2$:$52$:$43.0$.  The position of the BCG
nucleus appears to be $\sim 2^{\prime\prime}$ away from the peak in the X-ray surface brightness,
corresponding to one of the two bright spots in the cluster core \citep[see Figure 9 in ][]{2006McNamara}.

According to Figure 2 in \citet{2006McNamara}, the bulk of the  star formation, as measured from optical colours, occurs within $\simeq$ 10 kpc. In particular, the central blue colors associated with the starburst
are found at radii smaller than $7^{\prime\prime}$ (corresponding to $\sim 27$ kpc) from the BCG center
(see also their Figure 3).
Furthermore, recent ALMA observations \citep{2014McNamara} provide evidence of large amounts
of molecular gas at the level of $\simeq 5\times10^{10}$M$_\odot$ within a few kpc of the center.
However, the position of the molecular gas does not overlap with the position of the two X-ray bright regions,
where the gas is expected to cool most rapidly.
These two regions are partially included in the 10 kpc circle, and fully included in the 20 kpc circle.
Therefore, our choice of extracting two spectra in the inner 10 and 20 kpc nicely matches the central
starburst region and the maximum region encompassing all possible star formation events.

We find about 7500 and 26400  net counts (0.5-7 keV band)
in the  inner 10 and 20 kpc, respectively.  The background is sampled from the
ACIS-I CCD3 in a region far from the cluster.  Given the extent of this massive cluster,
the background may be contaminated by some emission from the cluster itself.  However, given the
small size of the regions under investigation, the background contribution is minimal and always below
0.1\% of the total emission.  We fit each region with our two methods (see \S 3.2 for details).
The redshift parameter is frozen; however, we verified that leaving it free does not have any noticeable
effect on the fit.  The Galactic hydrogen column density is frozen to the value
$N_{H} = 2.04 \times 10^{20}$ cm$^{-2}$, derived by \citet{2005LAB} through radio observations.

With our first model ({\tt mkcflow + mekal}) we measure a 95\% one-sided  confidence level,
i.e., upper limits (corresponding to $\Delta C_{stat} = 2.71$), of
$\dot M = 23 M_{\odot}$ yr$^{-1}$ in the inner 10 kpc
and  $22 M_{\odot}$ yr$^{-1}$ for 20 kpc (see Figure \ref{A1835mdot} and Table \ref{tab:spec_fits}).
The temperature of the {\tt mekal} component is rather stable, with best-fit values of $4.4\pm 0.2$
and $4.3\pm 0.1$ keV for the 10 and 20 kpc, respectively.
We find that in the 0.5-2 keV band, the contribution to the signal from gas below 3 keV is 6\% and   2\%
within 10 and 20 kpc, respectively.  This will be relevant in \S 5 when discussing possible
systematics associated with a global $\sim 3$\% uncertainty in the effective area of {\sl Chandra}.
We also postpone to \S 5 the discussion on  the effects of a larger $N_{H}$.


Next we fit our spectra with composite models consisting of five {\tt mkcflow} components in fixed temperature bins
and a {\tt mekal} component with free temperature (see \S 3.2 for details).  Our results for the two regions are
summarized in the two panels of Figure \ref{A1835mdot}, where the 95\% confidence level upper limit range from the single
{\tt mkcflow} model is shown by the shaded area (see also Table \ref{tab:spec_fits}).
As we can see, irrespective of the extraction radius, emission from gas below 1.8 keV is not
detected.  The lowest temperature range we investigate,
here and in other clusters, is characterized by a significantly less constraining upper limit
for the simple reason that emission from gas in this temperature range contributes only
in a modest way to the spectrum above 0.5 keV where we perform our measurements.
Above 1.8 keV, the gas seems to cool at a rate of $\sim 200-300 M_{\odot}$ yr$^{-1}$.
Our analysis suggests that the ICM in A1835  cools down to a temperature that is
approximately 2.4 keV, which is roughly $1\over 4$ of the ambient temperature, $9.7\pm 0.14$ keV, measured from the
circular ring defined by 40 kpc $< r < $400 kpc.
Below this temperature, there is no clear evidence of significant cooling.
If we require the four {\tt mkcflow} components covering the 0.15-1.8 keV  range to have the same mass deposition
rate, we find that $\dot M \leqslant 17.5 M_\odot $ yr$^{-1}$ and $10 M_\odot $ yr$^{-1}$
at a $95\%$  confidence level (corresponding to
$\Delta C_{stat} = 2.71$) for the 10 and 20 kpc  regions, respectively.
We conclude that the spectral mass cooling rate $\dot M$ in the starburst region is lower by about one order of magnitude than the star formation rate in the BCG (see Table 1).

Finally, we compare our results based on low-resolution spectra with previous results
based on grating spectra and published in the literature \citep{2003Peterson,2010Sanders}.
The advantage of RGS spectra over CCD spectra is that it affords direct detection of
emission lines associated with plasma of different temperatures.  As pointed out by
\citet{2010Sanders}, the RGS spectrum of A1835 shows no evidence of emission lines from plasma
with temperatures below $\sim 3$ keV. Under these circumstances, the breakup of the spectrum in different
components relies mostly on its overall shape, and RGS data is no better suited for this purpose
than CCD data.  Actually, under the assumption that the mass cooling occurs
in the same region where the star formation is taking place, i.e., roughly within
$20$ kpc of the center \citep{2006McNamara}, the limited  spatial resolution of the
RGS, $\sim  30$ arcsec corresponding to $\sim  120$ kpc, results in a
significant disadvantage as any emission from cool thermal components coming from within
$\sim 30$ kpc will be swamped by the emission from the much larger RGS extraction region.
This is likely the reason for the order of magnitude large upper limit to the mass cooling rate,
$\dot M < 200 M_{\odot}$ yr$^{-1}$ \citep[][]{2010Sanders} and $\dot M < 140 M_{\odot}$ yr$^{-1}$ \citep[][]{2003Peterson},
both at a 90\% confidence level, found in A1835 with XMM-Newton gratings data.

\subsection{RXCJ1504}

\begin{figure*}
\includegraphics[width=8cm]{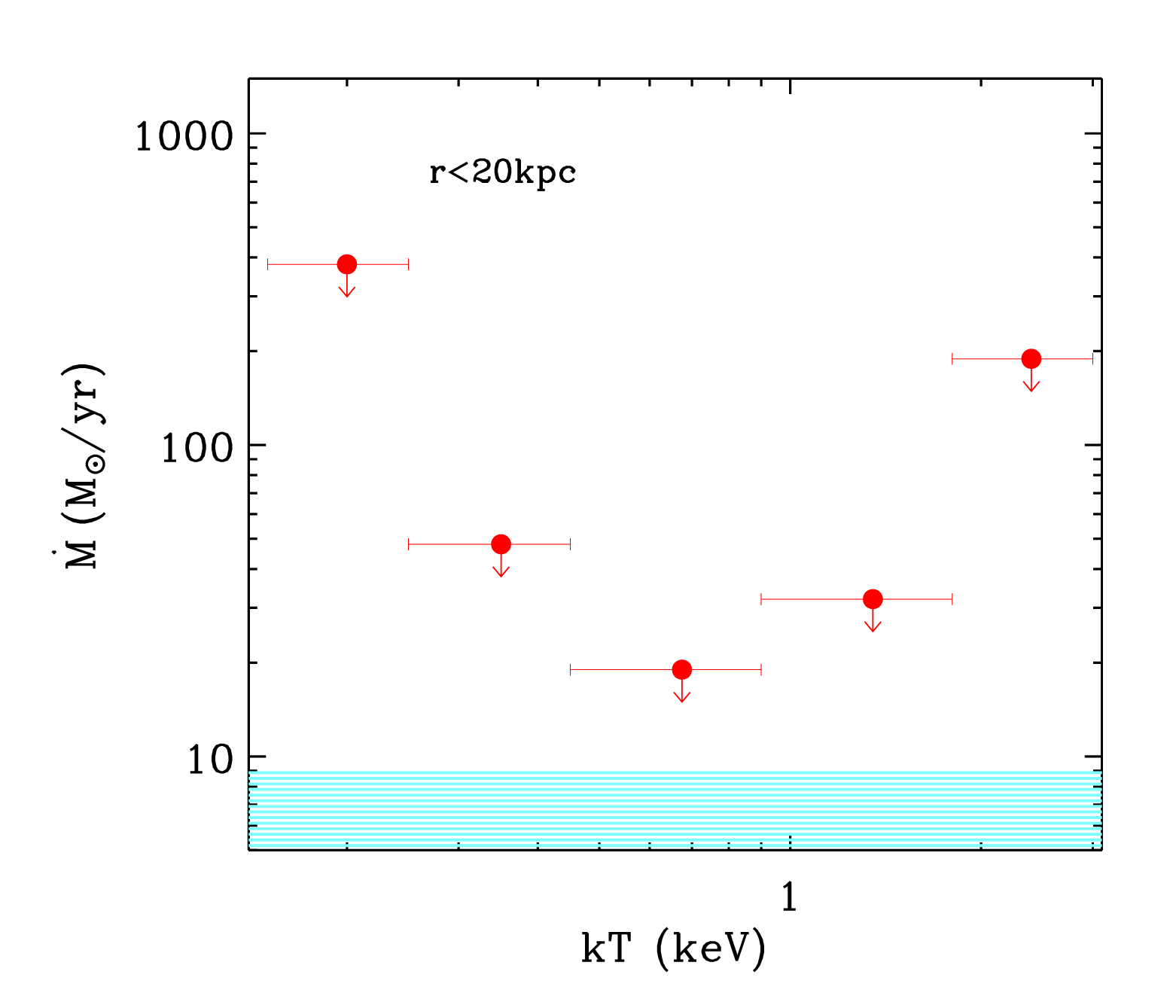}
\includegraphics[width=8cm]{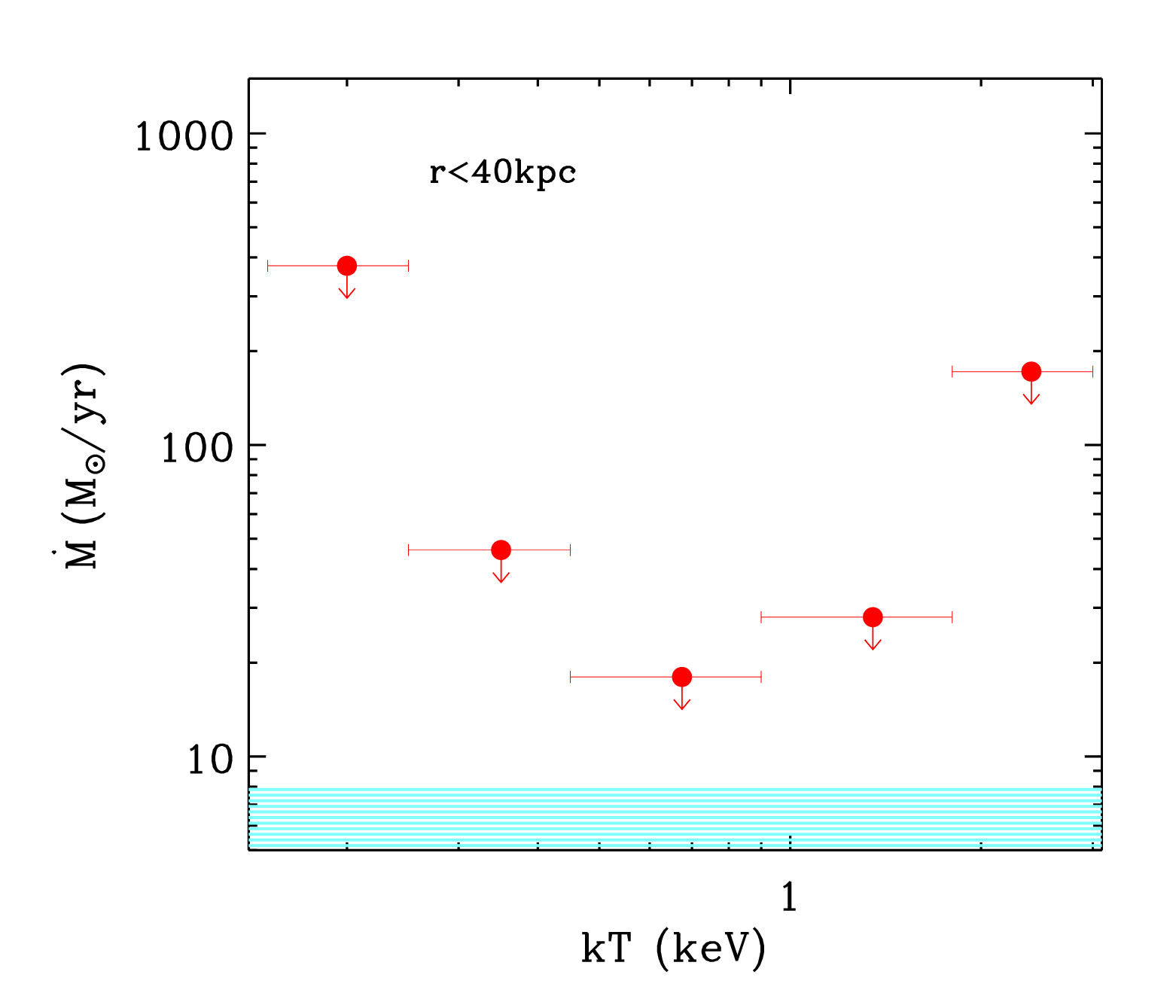}
\caption{\label{RXCJ1504mdot} Measured cooling rate in RXCJ1504 in the inner 20 kpc (left panel)
and 40 kpc (right panel). Symbols  as in Figure \ref{A1835mdot}.}
\end{figure*}

RXCJ1504 was observed with ACIS-I for a total of  13.2 ks of effective exposure
in the FAINT mode and for 39 ks in the VFAINT mode.
We measure the redshift of the cluster from the global X-ray spectrum finding
$z_X=0.2126_{-0.0023}^{+0.0029}$, in agreement, within $1 \sigma$, with the optical value
$z_{opt}=0.2153$ \citep{2010Ogrean}.  We prefer to use the value of $z_X$ which makes the
fit more sensitive to the presence of line blends from cool gas.
At this redshift the angular diameter scale is $3.45$ kpc/arcsec.  Therefore
20 kpc corresponds to 11.6 physical pixels and  40 kpc
to 23.2 pixels.   We choose the center to be on the center of the BCG
galaxy, $RA_{BCG}=15$:$04$:$07.490$, $DEC_{BCG}=-2$:$48$:$15.97$,
which is within $\sim 5 $ kpc from the peak in the X-ray surface brightness.
Most star formation occurs in the core of the BCG and in a 42-kpc-long filament of
blue continuum, line emission and X-ray emission, that extends southwest of the galaxy and that
is entirely included in a circle with a 40 kpc radius centered on the BCG \citep[see Fig. 9 in][]{2010Ogrean}.
Therefore, for this cluster we perform our analysis in two circles  with radii of 20 and 40 kpc.

We find  $\sim$  16600 ($5700$)  and $39400$ ($13600$)
net counts ($0.5$-$7$ keV band) within  20 and 40 kpc of the center of the BCG, respectively,
in the ACIS-I VFAINT (FAINT) observation.
The background is sampled from the ACIS-I CCD3, in a region far from the cluster.  Also in this case,
given the extent of this nearby massive cluster,
the background is contaminated by some emission from the cluster itself.  However, given the
small size of the region we are  investigating, the background contribution is minimal and always below
0.2 \% of the total emission.  The value of the Galactic hydrogen column density is
$N_{H} = 5.94 \times 10^{20}$ cm$^{-2}$ according to \citet{2005LAB}.

With the single {\tt mkcflow} model in the 0.15-3.0 keV temperature range, we find
very low 95\% confidence level upper limits on the order of  10 $M_\odot$ yr$^{-1}$
irrespective of the extraction radius (see Fig. \ref{RXCJ1504mdot} and Table \ref{tab:spec_fits}).
The results from the multi-component models, show that the two lowest temperature bins (energies
below 0.5 keV) are almost unconstrained owing to the large  Galactic absorption, which removes a
large part of the signal at the lowest energies.  As shown in Figure \ref{RXCJ1504mdot} (see also Table \ref{tab:spec_fits})
the $95\%$ upper limit on $\dot M$ in the range $0.5$-$1.8$ keV is $\sim 30 \, M_\odot$ yr$^{-1}$,
irrespective of the extraction radius.  It grows to $\sim 180 M_\odot$ yr$^{-1}$
in the temperature range $1.8-3.0$ keV.
If we require the four {\tt mkcflow} components covering the 0.15-1.8 keV  range to have the same mass deposition
rate, we find that $\dot M \leqslant 10 M_\odot$ yr$^{-1}$ and $\dot M \leqslant 8 M_\odot$ yr$^{-1}$  at
a $95\%$  confidence level (corresponding to $\Delta C_{stat} = 2.71$)
for the 20 and 40 kpc  regions, respectively, which are, within the errors, consistent with those from
the single {\tt mkcflow} model.
These values are more than an order of magnitude smaller than the SFR in the BCG.
Our results can be compared to those reported by \citet{2010Ogrean}, who find $3\sigma$ upper limits of
156$\, M_\odot$ yr$^{-1}$ and 239$\, M_\odot$ yr$^{-1}$ from the analysis of EPIC and RGS spectra respectively.
As in the case of A1835 the far less constraining XMM-Newton limits likely follow from the
much larger extraction regions chosen by these authors: namely 0.67 arcminutes (140 kpc)
for EPIC and 2 arcminutes (414 kpc) for RGS.


\subsection{RXJ1532}

\begin{figure*}
\includegraphics[width=8cm]{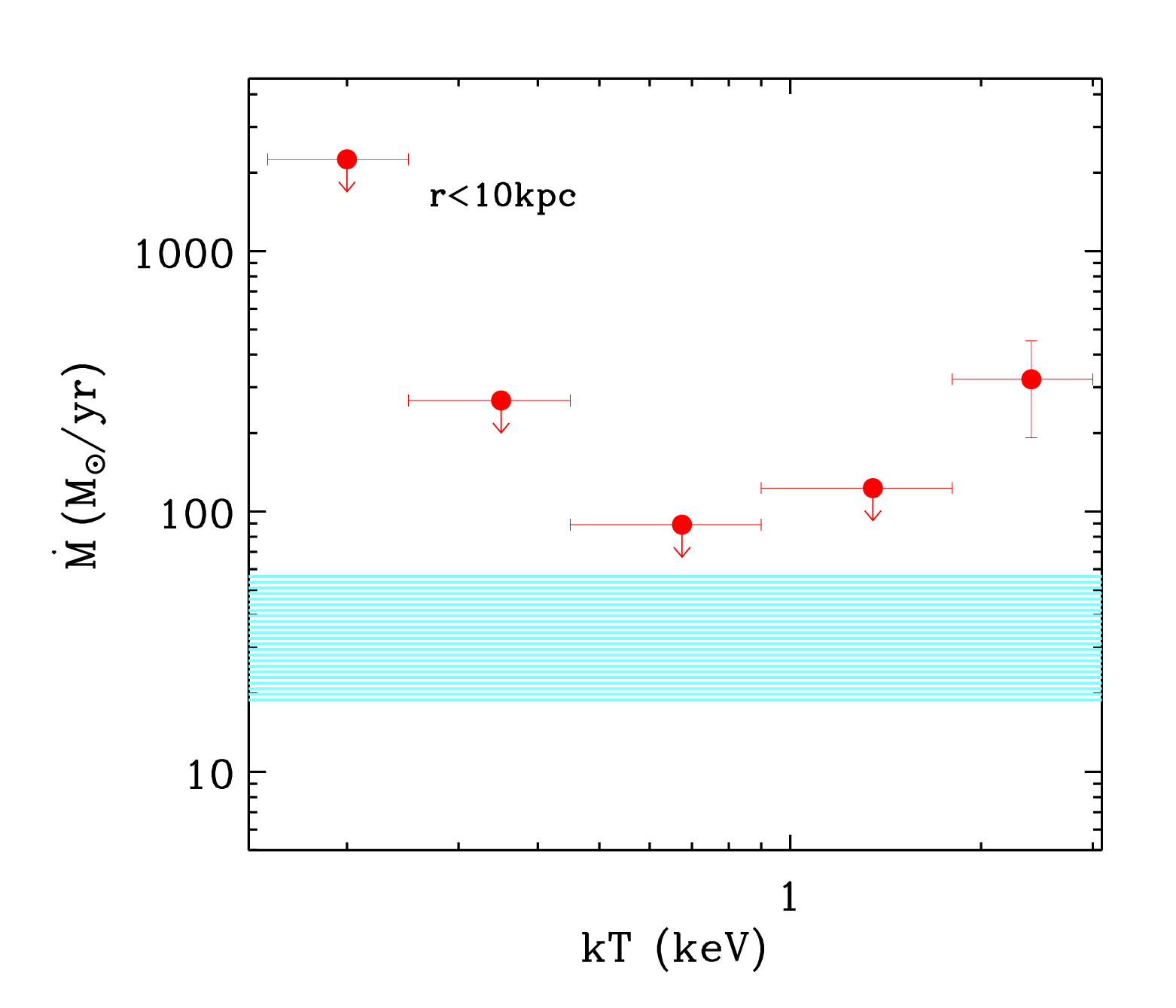}
\includegraphics[width=8cm]{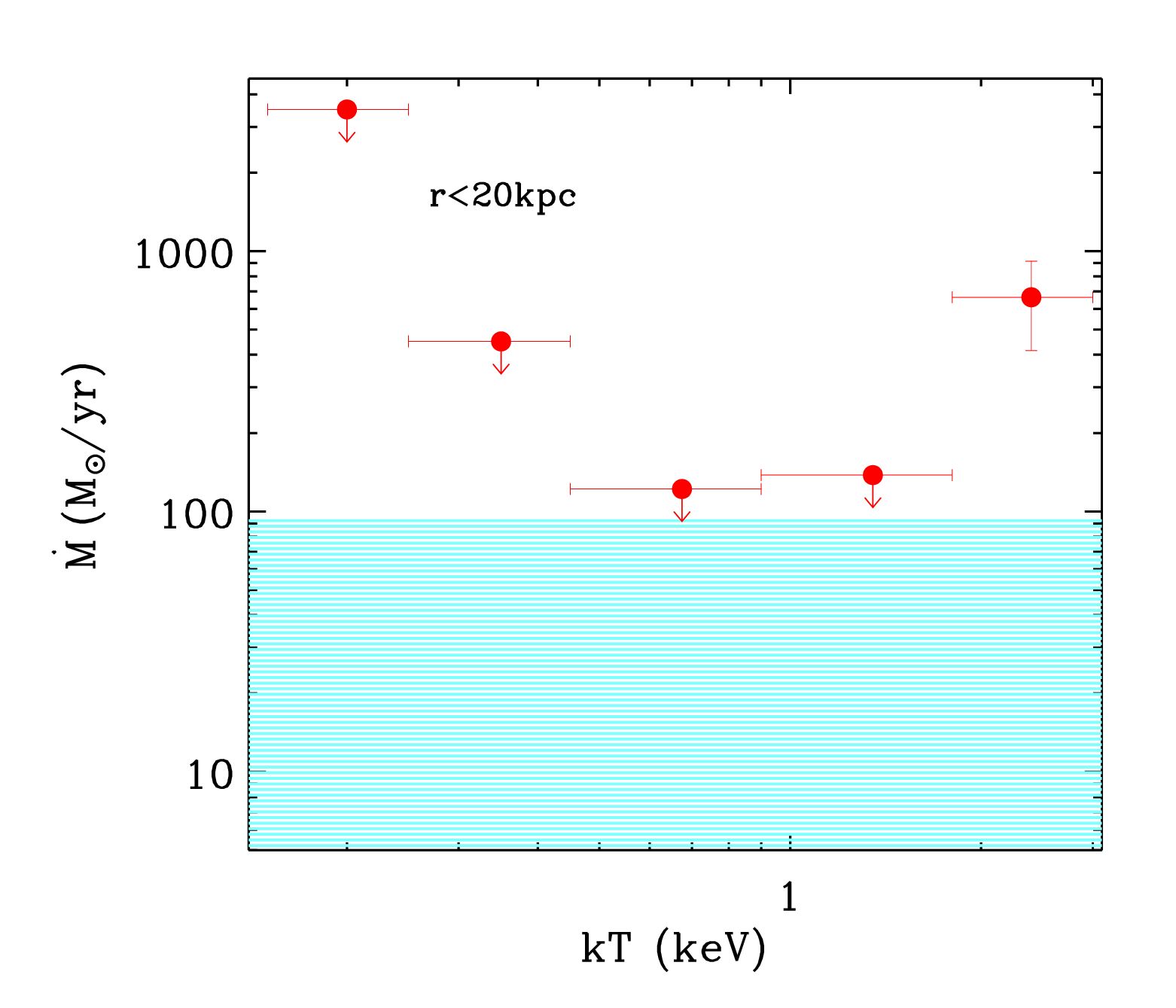}
\caption{\label{RXJ1532mdot}Measured cooling rate in RXCJ1532
in the inner 10 kpc (left panel) and 20 kpc (right panel).
Symbols as in Figure \ref{A1835mdot}. }
\end{figure*}

RXJ1532 was observed  with ACIS-I and ACIS-S in 2001 for about 10 ks for both exposures.
However, in 2011, a much deeper exposure of 88 effective ks with ACIS-S, VFAINT mode,
has been acquired.  Therefore, we use this last ACIS-S observation for our analysis.
We measure the redshift of the cluster from the global X-ray spectrum finding
$z_X=0.361_{-0.0015}^{+0.0014}$, this is the value that will be used in the X-ray analysis.
We verified that when the redshift is left free during the fits,
its best-fit value does not change significantly.
The star formation observed in the BCG has been measured to be $110 M_\odot $ yr$^{-1}$
by \citet{2012Hoffer}.  Recently, \citet{2013Hlavacek-Larrondo}, using HST data, has derived the total SFR
associated with the BCG within a circular region centered on the galaxy with a radius of 3.5 arcsec ($\sim 18$
kpc), while beyond this region they detected no significant UV emission\footnote{The UV SFR measured
by \citet{2013Hlavacek-Larrondo} is  $76\pm38$ M$_\odot$/yr, consistent with the IR measure by \citet{2012Hoffer}.}.
Therefore, we choose to extract spectra from two circular regions of 10 and 20 kpc in radius centered
on the BCG at $RA_{BCG} = 15$:$32$:$53.796$, $DEC_{BCG}=+30$:$20$:$58.97$.

The angular diameter scale for RXJ1532 is $5.04$ kpc/arcsec,
10 kpc corresponds to 4 physical pixels on ACIS-S and  20 kpc to 8 pixels.
We find  2500,  7900  net counts (0.5-7 keV band) within 10 and 20 kpc of the center, respectively.
The background is sampled from the ACIS-S CCD7, and as for our other systems, the background contribution
is low, always below 0.2 \% of the total emission.

The value of the Galactic hydrogen column density is
$N_{H} = 2.3 \times 10^{20}$ cm$^{-2}$ according to \citet{2005LAB}.
For the single {\tt mkcflow} model we have a marginal detection (about $2\sigma$)  of a non-negligible
mass cooling rate within 10 kpc, $38 \pm 20  M_\odot $ yr$^{-1}$, while for 20 kpc we derive an upper
limit of about 100$M_\odot $ yr$^{-1}$ (see Fig. \ref{RXJ1532mdot} and Table \ref{tab:spec_fits}).
 The temperature of the {\tt mekal}
component is $kT = 4.5\pm 0.4$ and $4.2\pm 0.2$ keV for 10 and 20 kpc, respectively.
The multi-component analysis shows, as for A1835, no evidence of emission below 1.8 keV
(see Fig. \ref{RXJ1532mdot} and Table \ref{tab:spec_fits}).
If we require the four {\tt mkcflow} components covering the 0.15-1.8 keV  range to have the same mass deposition
rate, we find that $\dot M \leqslant 49  M_\odot $ yr$^{-1}$  at a 95\% confidence level, corresponding to $\Delta C_{stat} = 2.71$,
for the 10 kpc region. A somewhat less constraining upper limit, $\dot M \leqslant 67 M_\odot $ yr$^{-1}$, is found
for the 20 kpc region that presumably encompasses the entire star formation region.
We conclude that the currently available data is of insufficient statistical quality to provide
constraints for a useful comparison with the SFR.



\subsection{Abell 1068}

\begin{figure*}
\includegraphics[width=8cm]{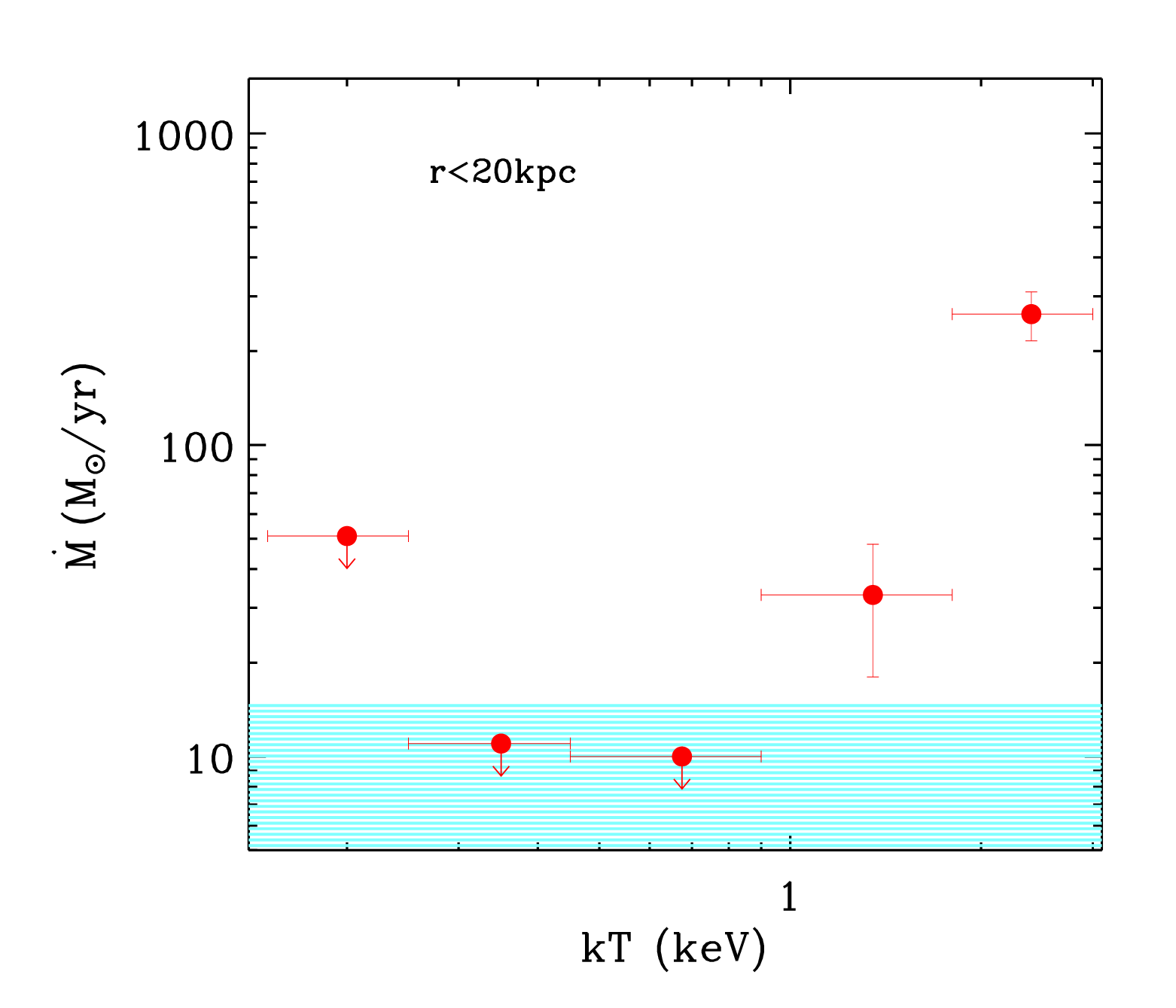}
\includegraphics[width=8cm]{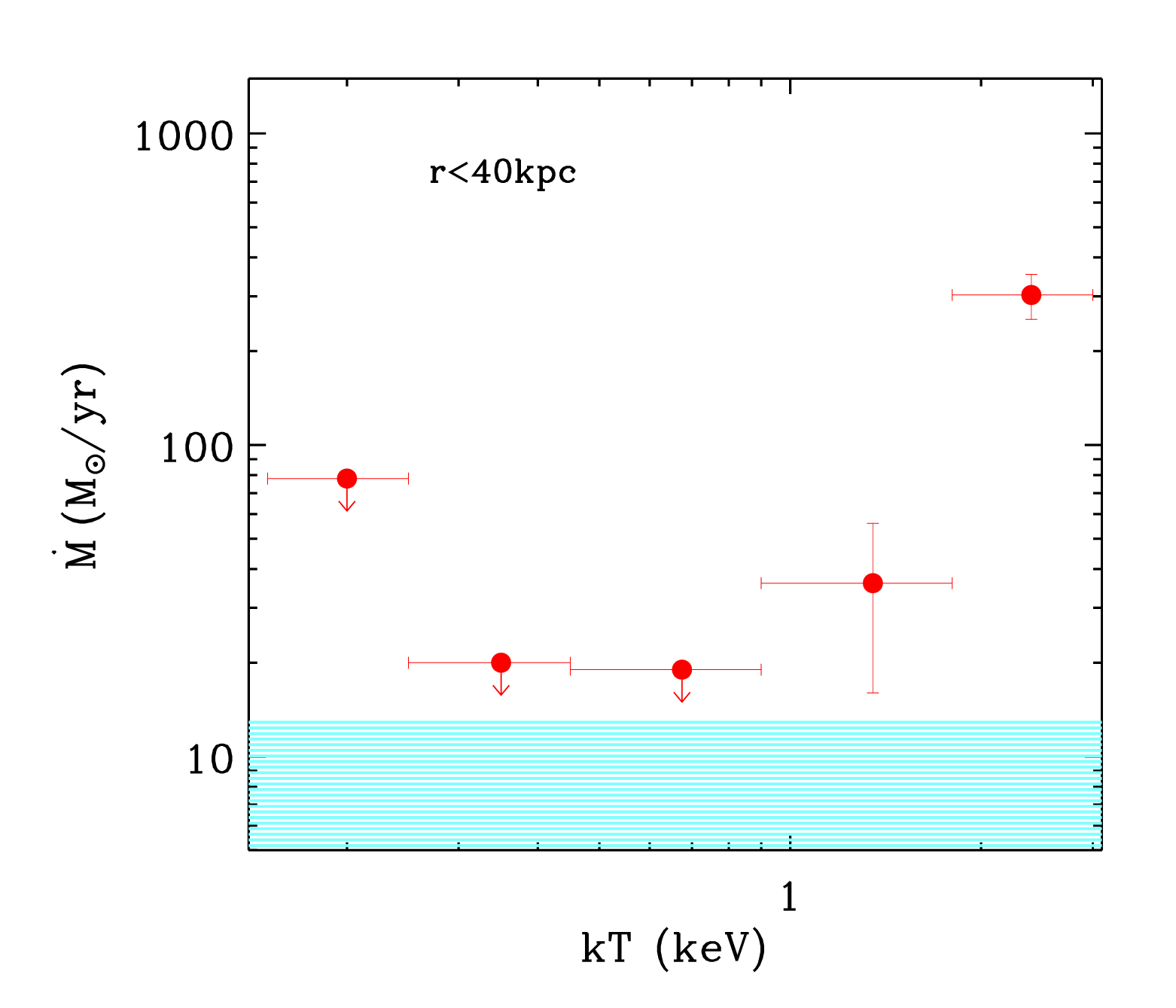}
\caption{\label{A1068mdot}Measured cooling rate in A1068
in the inner 20 kpc (left panel) and 40 kpc (right panel).
Symbols as in Figure \ref{A1835mdot}. }
\end{figure*}

The {\sl Chandra} data on A1068 consist of 30 ks with ACIS-S in the FAINT mode.
The total exposure time amounts to 26.6 ks after the standard
data reduction procedure.  A1068 is also observed in other ACIS-I and ACIS-S
pointings, but always at large off-axis angles, therefore
we do not consider further data for our analysis.  The spectral analysis of the ACIS-S data
published in \citet{2004Wise}
found a total mass cooling rate of $\sim 150 M_\odot$ yr$^{-1}$ from the total emission,
while only $\sim 40 M_\odot$ yr$^{-1}$ within the inner 30 kpc.
In the same paper the authors discuss the  possibility that the observed starburst may  also be triggered by
interactions between the BCG and a group of companion galaxies rather than by a
cooling flow.  According to \citet[][see their Figure 2]{2004Wise} the bulk of the starburst is contained
within $\simeq 4^{\prime\prime}$ ($\sim 10$ kpc) of the center and is fully contained within
$\simeq 6^{\prime\prime}$ (15 kpc).  A secondary starburst is located approximately
15$^{\prime\prime}$ (35 kpc) SE of the core, and one anomalous blue region  is seen up to 35$^{\prime\prime}$ from the optical center (see their Figure 3).  Finally, according to Figure 1 in \citet{2004Wise}, the optical peak
is offset west with respect to the X-ray peak by about 3$^{\prime\prime}$.
Conversely, according to \citet{2010Edge} the IR emission is on scales $< 5^{\prime\prime}$.
To include all possible cooling sites, we center our extraction regions on the optical position of the BCG
($RA_{BCG}= 0$:$40$:$44.508$; $DEC_{BCG} =+39$:$57$:$11.23 $) and consider extraction radii of 20 and 40 kpc.

At the cluster redshift the angular diameter scale is $2.44$ kpc/arcsec.  Therefore, on ACIS-S,
20 kpc corresponds to 16.7 physical pixels  and 40 kpc
to 33.3 pixels.  We fix the redshift to the value obtained from the best fit to the total spectrum
$z_X =  0.1374_{-0.003}^{+0.001}$.
We identify the presence of a weak AGN in the hard 2-7 keV image, and therefore
remove the inner 2.2 kpc to exclude its emission.
We find   9160 and 17500 net counts (0.5-7 keV band) within
20 and 40 kpc, respectively.  The background is sampled from the
ACIS-S CCD7 as in the previous cases, and the background contribution
is always below 0.2 \% of the total emission.  The value of the Galactic hydrogen column density is
$N_{H} = 1.69 \times 10^{20}$ cm$^{-2}$ according to \citet{2005LAB}.

For the single {\tt mkcflow} model we obtain  $\dot M <15$ $M_{\odot}$ yr$^{-1}$ and
$\dot M <13$  $M_{\odot}$ yr$^{-1}$ for  20 and 40 kpc, respectively (see Fig. \ref{A1068mdot} and Table \ref{tab:spec_fits}).
The multi-component {\tt mkcflow} model highlights tight upper limits on emission from gas with temperature in the range 0.25-0.9 keV (see Fig. \ref{A1068mdot} and Table \ref{tab:spec_fits}). Emission could be present in the energy range 0.9 to  1.8 keV, albeit at a low statistical significance of $ \sim 2\sigma$.
If we require the four {\tt mkcflow} components covering the 0.15-1.8 keV  range to have the same mass deposition
rate, we find that $\dot M \leqslant 10 M_\odot $ yr$^{-1}$ and $\dot M \leqslant 5  M_\odot $ yr$^{-1}$  at a 95\%  confidence level, corresponding to $\Delta C_{stat} = 2.71$ for the 20 and 40 kpc region, respectively.
Thus, for A1068, our upper limit is lower by an order of magnitude or more with respect to the SFR reported by \citet{2012Rawle}.

\subsection{ZW3146}

\begin{figure*}
\includegraphics[width=8cm]{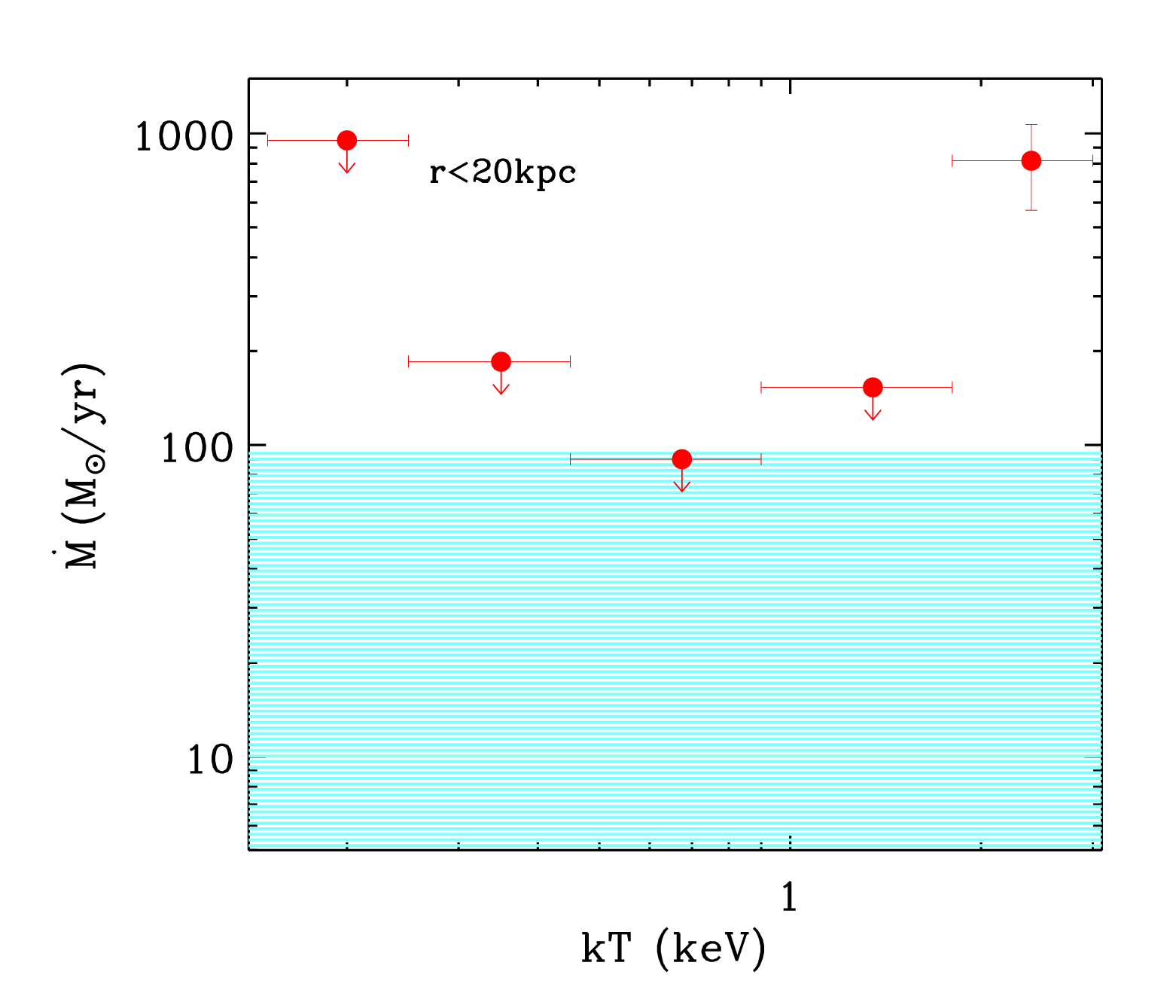}
\includegraphics[width=8cm]{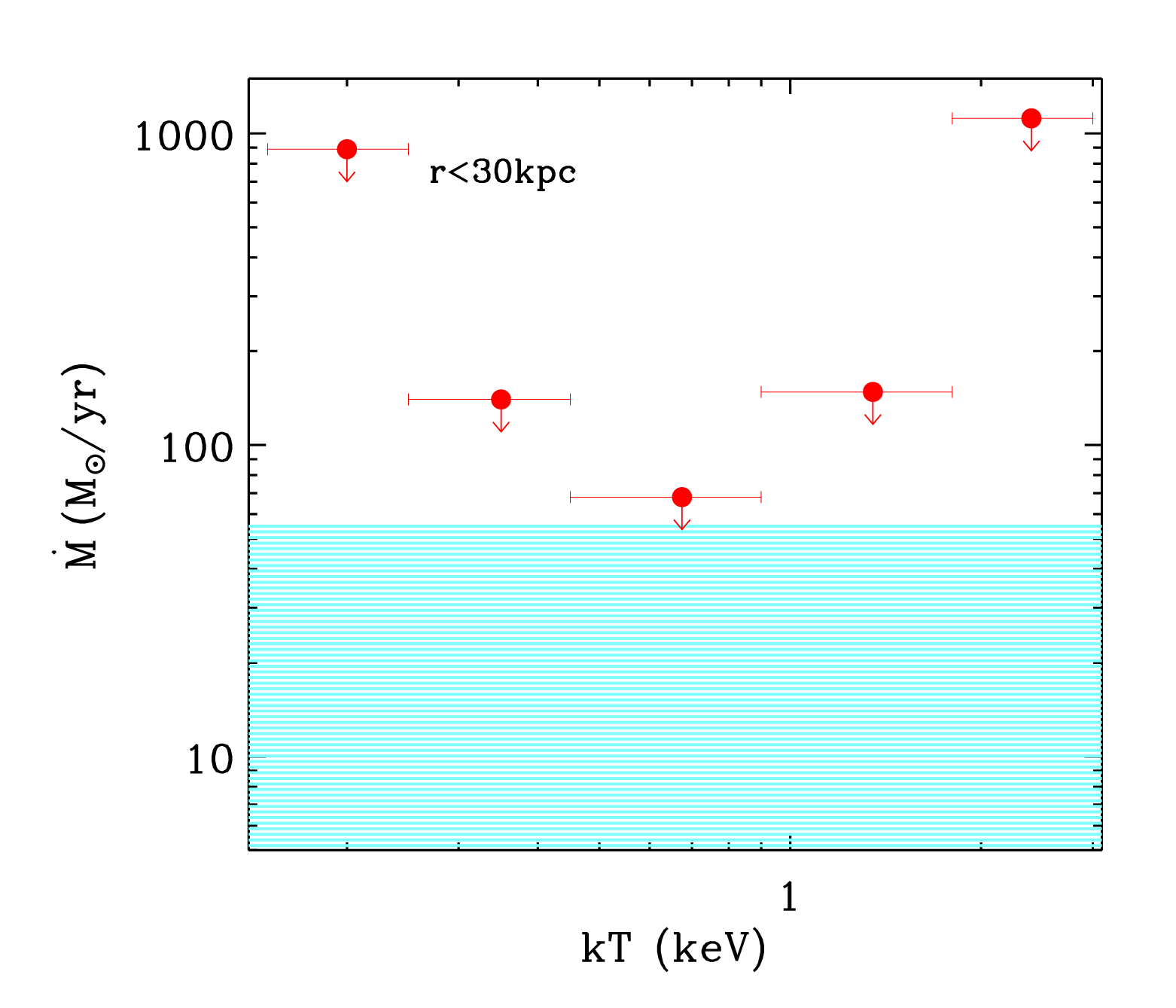}
\caption{\label{ZW3146mdot}Measured cooling rate in ZW3146
in the inner 20 kpc (left panel)  and 30 kpc (right panel).
Symbols as in Figure \ref{A1835mdot}. }
\end{figure*}

ZW3146 was observed with ACIS-I for a total of  45.6 ks of effective exposure
in the FAINT mode and for 40 ks in the VFAINT mode.
We measure the redshift of the cluster from the global X-ray spectrum finding
$z_X=0.2897_{-0.0029}^{+0.0026}$, and we keep this value frozen in the spectral fits.
At this redshift the angular diameter scale is $4.36$ kpc/arcsec.  Therefore,
20 kpc corresponds to 9.3 physical pixels on ACIS-I and 30 kpc to 14 pixels.
ZW3146 hosts a significant amount of star formation in the BCG.  According to Table 4 in \citet{1999Crawford}
the $H_\alpha$ flux comes from a slit with a
size of $\sim 6^{\prime\prime}$ corresponding to a width of 26 kpc. Moreover, according to \citet{2010Edge},
the dust emission has an extent comparable to the bulk of the CO emitting gas and optical
emission lines (less than $5^{\prime\prime}$).  Finally, the X-ray center is offset  from the peak of UV starburst
\citep[see][]{2010Odea}.  Therefore, we choose the center to be on the center of the BCG
galaxy, $RA_{BCG}=10$:$23$:$39.6$, $DEC_{BCG}=+04$:$11$:$10.8$,
which is within $\sim 10 $ kpc of the centroid of the X-ray surface brightness.
We  perform extractions within 20 and 30 kpc radii to encompass any possible region with significant star formation.

We find  $\sim 4140$ ($4650$)  and $6820$ ($7800$)
net counts ($0.5-7$ keV band) within 20 and 30 kpc, respectively, in the ACIS-I VFAINT (FAINT) observation.
The background is sampled from the ACIS-I CCD3 in a region far from the cluster core.  Also in this case,
given the extent of this nearby and massive system,
the background is contaminated by some outer emission of the cluster itself.  However, given the
small size of the region we investigate, the background contribution is minimal and always below
0.1\% of the total emission.  The value of the Galactic hydrogen column density is
$N_{H} = 2.46 \times 10^{20}$ cm$^{-2}$ according to \citet{2005LAB}.

With the single {\tt mkcflow} model in the 0.15-3.0 keV temperature range, we find
$\dot M  < 94\, M_\odot$ yr$^{-1}$ and $\dot M  < 56 \, M_\odot$ yr$^{-1}$ for  20 and 30 kpc, respectively (see Fig. \ref{ZW3146mdot}
and Table \ref{tab:spec_fits}).
The temperature of the {\tt mekal} component is around 4 keV ( $4.1\pm 0.2$ and  $4.2\pm 0.1$
keV for  20 and 30 kpc, respectively).
From the results obtained with the multi-component {\tt mkcflow} models, shown in Figure \ref{ZW3146mdot} and Table \ref{tab:spec_fits},
we find that the tightest constraints on emission come from the 0.45-0.9 keV temperature interval.
If we link the four {\tt mkcflow} components covering the range  0.15-1.8 keV we find
an upper limit of  $\dot M \leqslant 45 M_\odot $ yr$^{-1}$ and $\dot M \leqslant 34 M_\odot $ yr$^{-1}$  at a 95\%  confidence level, corresponding to $\Delta C_{stat} = 2.71$ for the 20 and 30 kpc region, respectively.
Despite these rather loose constraints, the cooling rates are a factor of 2-3 smaller than the measured star formation rate
(see Table \ref{tab:zobs}).

A previous estimate of the spectroscopic mass cooling rate was performed in Rafferty et al. (2006),
using Chandra ACIS spectra; however, the extraction region was significantly larger, namely 186 kpc.
These authors found a cooling rate of  $\dot M \leqslant 590^{+190}_{-170}  M_\odot $ yr$^{-1}$.


\subsection{Z0348}

\begin{figure*}
\includegraphics[width=8cm]{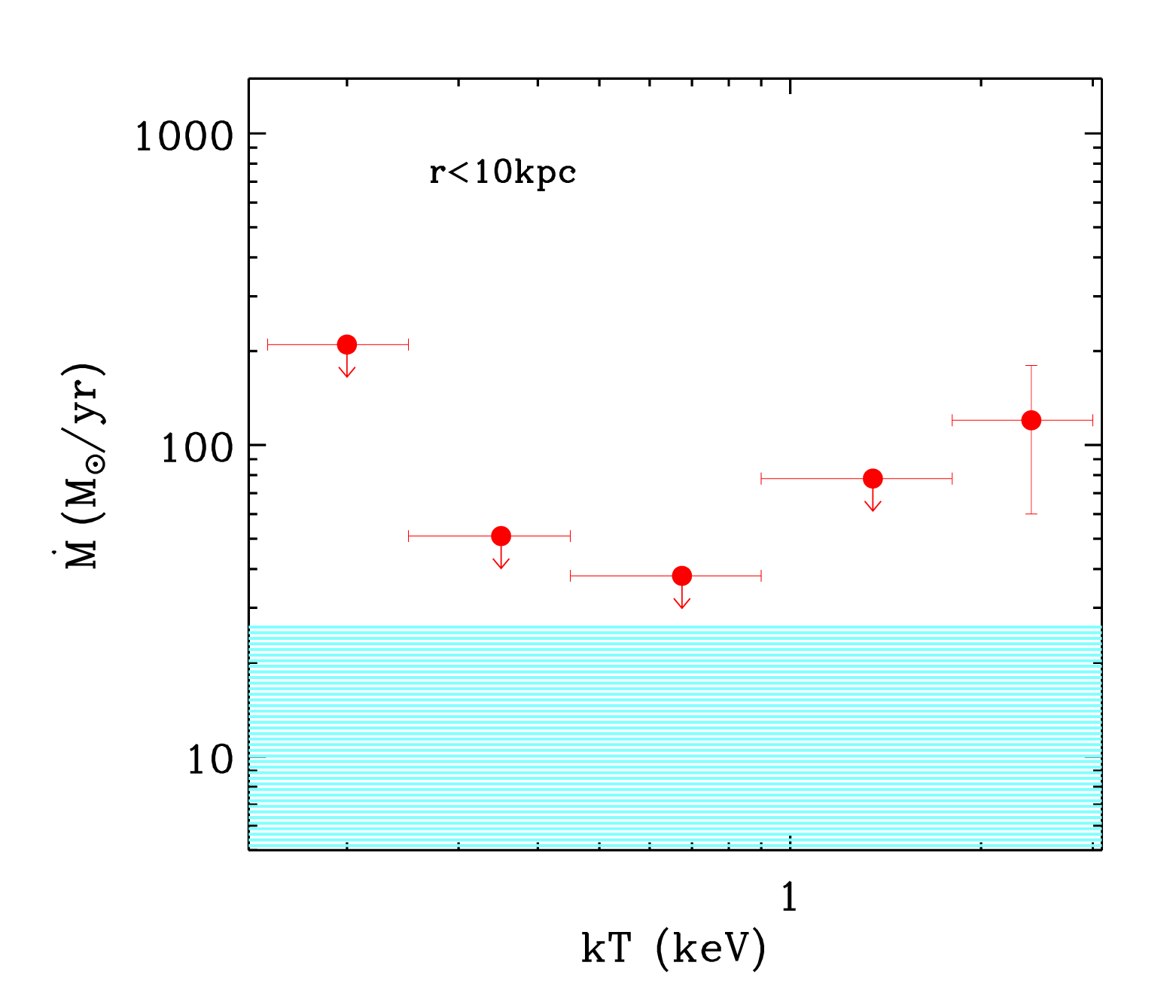}
\includegraphics[width=8cm]{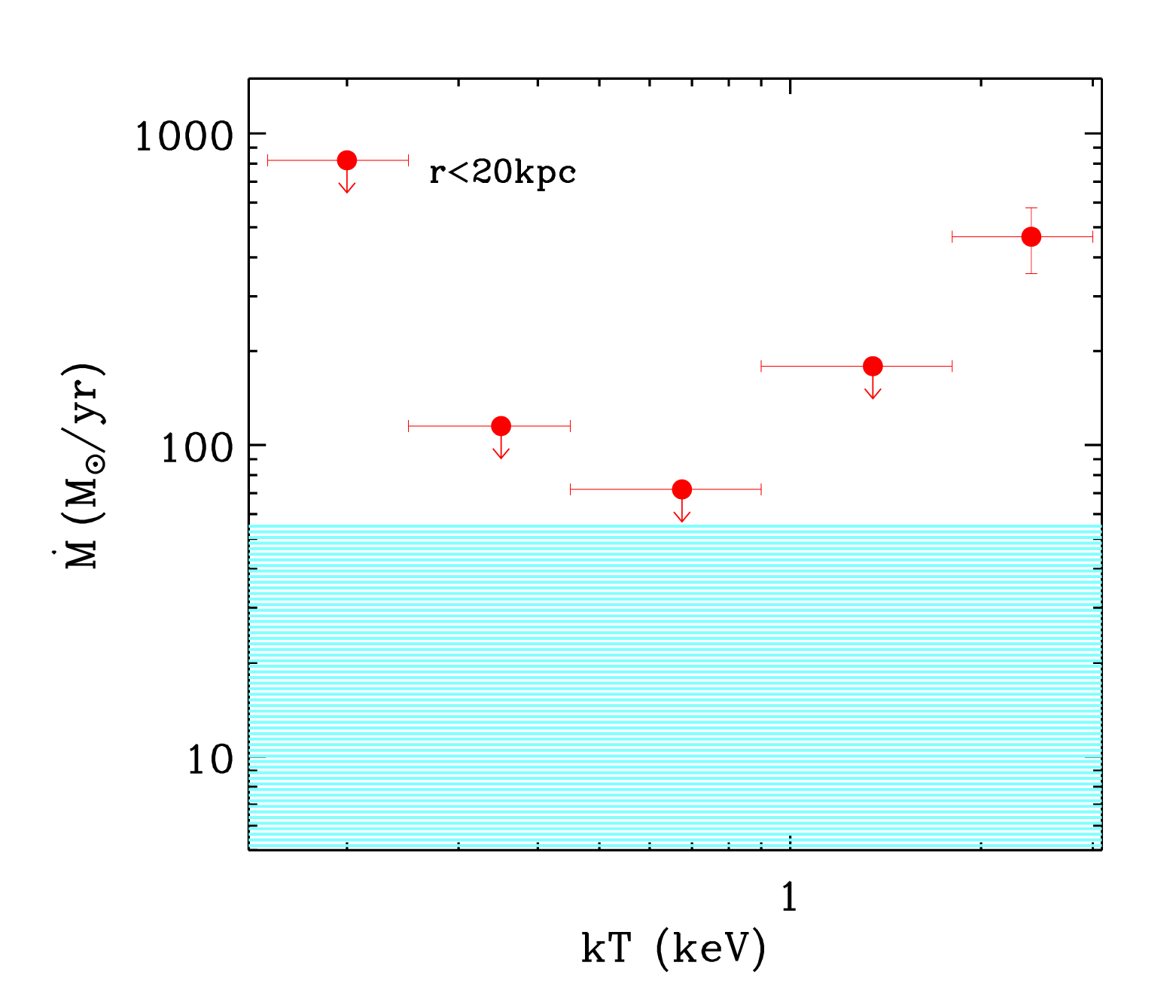}
\caption{\label{Z0348mdot}Measured cooling rate in Z0348
in the inner 10 kpc (left panel) and 20 kpc (right panel).
Symbols as in Figure \ref{A1835mdot}.  }
\end{figure*}

The {\sl Chandra} data on Z0348 consist of 47.8  ks, after the standard
data reduction procedure,  with ACIS-S in the VFAINT mode.
We detect an AGN in the center of the cluster, and we remove it
for the sake of our spectral analysis. As in the case of the Phoenix cluster \citep{2012McDonald}
the  X-ray emission of this active nucleus in the middle of a bright cool core appears to be strongly absorbed.
We fix the redshift to the value obtained from the best fit to the total spectrum,
which gives $z_X = 0.256 \pm 0.002$, in very good  agreement with the optical
value $z=0.254$.  At the cluster redshift the angular diameter scale is $3.97$ kpc/arcsec.
Therefore, 10 kpc corresponds to $5.1$ physical pixels and 20 kpc
to $10.2$ pixels.
We choose the center to be on the X-ray peak, which coincides with the position of the AGN and
with the center of the BCG at $RA_{BCG}=1$:$06$:$49.410$, $DEC_{BCG}=+1$:$03$:$22.55$
\citep[see][]{2010Odea}.  The BCG in Z0348 features  diffuse Ly$_\alpha$
emission extending  out to some 5$^{\prime\prime}$ (20 kpc) from the position
of the unresolved VLA radio source, coincident with the X-ray AGN.  Beyond this value, the Ly$_\alpha$
emission is negligible \citep{2010Odea}. We extract spectra from two regions with bounding radii
of 10 kpc and 20 kpc, encompassing all star forming sites.

We measure  1100 and  4150 net counts (0.5-7 keV band) within 10 and 20 kpc, respectively,
after removing  the inner 2$^{\prime\prime}$, contaminated by the AGN.  The background is sampled from the
ACIS-S CCD7, as it was in previous cases, and the background correction
is always below 0.1\% of the total emission.  The value of the Galactic hydrogen column density is
$N_{H} = 2.5 \times 10^{20}$ cm$^{-2}$ according to \citet{2005LAB}.

For the single {\tt mkcflow} model we obtain an upper limit of
$\dot M < 26.4  $ $M_{\odot}$ yr$^{-1}$ and $ \dot M < 56.1 $  $M_{\odot}$ yr$^{-1}$
for 10 and 20 kpc, respectively (see Figure \ref{Z0348mdot}
and Table \ref{tab:spec_fits}).
We note that for Z0348 the ambient temperature is rather low,
leaving little room to constrain the mass cooling rate.

The analysis with the multi-component models shown in Figure \ref{Z0348mdot} and Table \ref{tab:spec_fits}
confirms that the upper limits allow for a significant cooling rate.
If we consider the 0.45-0.95 keV range where constraints are tightest,
 we find upper limits of $38 M_{\odot}$ yr$^{-1}$ and $ 72 M_{\odot}$ yr$^{-1}$ at a 95\% confidence level for 10 and 20 kpc, respectively.
By linking the four {\tt mkcflow} components covering the range 0.15-1.8 keV,
we find upper limits of $21.3 M_{\odot}$ yr$^{-1}$ and $ 47.9 M_{\odot}$ yr$^{-1}$ for 10 and 20 kpc, respectively.
These values are broadly consistent with the SFR in the BCG.  We conclude that in Z0348
there is no substantial disagreement between star formation and mass cooling rate.
A deeper exposure on this cluster should allow us to put stronger constraints on the mass cooling rate.

\begin{table*}
\centering
\caption{\label{tab:spec_fits}Best-fit values for $\dot M$ obtained from single and multi {\tt mkcflow}
models. Upper limits are at the 95\% confidence level while errors are 1$\sigma$.}
\begin{tabular}{|c|c|c|c|c|c|c|c|}
\hline

Name     &   region   &  single {\tt mkcflow}  &multi {\tt mkcflow}&multi {\tt mkcflow} &multi {\tt mkcflow}&multi {\tt mkcflow} & multi {\tt mkcflow}  \\
         &            &                        &  0.15-0.25 keV    & 0.25-0.45 keV      & 0.45-0.9 keV      & 0.9-1.8  keV       & 1.8-3.0  keV         \\
         &     kpc    &    $M_\odot$yr$^{-1}$  &$M_\odot$yr$^{-1}$ &$M_\odot$yr$^{-1}$  &$M_\odot$yr$^{-1}$ & $M_\odot$yr$^{-1}$ &     $M_\odot$yr$^{-1}$  \\
\hline
  A1835  &      10    &   23.3                 &  190              & 34                 &        28         &      73            & $167^{+40}_{-60}$    \\
         &      20    &   21.7                 &  209              & 36                 &        21         &      41            & $367\pm 176  $       \\
\hline
RXCJ1504.1 & 20       &    10.0                 &  380              & 48                 &        19         &      32            & 189                  \\
           & 40       &    8.0                 &  376              & 46                 &        18         &      28            & 172                  \\
\hline
RX J1532.9+3021 & 10  & $ 38.2 \pm 20$         &  2253             & 267                &        89         &     123            & $322 \pm 130$        \\
                & 20  &   95.0                 &  3500             & 450                &       122         &     138            & $665 \pm 250$        \\
\hline
A1068           & 20  &   15                   &    51             &  11                &        10         &  $33 \pm 15$       & $263 \pm 47$         \\
                        & 40  &   13                   &    78             &  20                &        19         &  $36 \pm 20$       & $303 \pm 50$         \\
\hline
ZW3146      & 20          &   94.1                 &   950             & 185                &        90         &     153            & $817.2 \pm 250$      \\
            & 30      &   55.8                 &   890             & 140                &        68         &     148            & 1118                 \\
\hline
Z0348       & 10      &   26.4                 &   210             &  51                &        38         &      78            & $120 \pm 60$         \\
            & 20          &   56.1                 &   820             & 115                &        72         &     179            & $466 \pm 111$        \\
\hline
\end{tabular}
\end{table*}

\subsection{Summary of measures on individual systems}

For all our systems, once emission above a given threshold is excluded, we find  no significant evidence of cooling gas. The threshold is found to be 1.8 keV for all systems but  A1068 and RXJ1504, whose emission is detected down to 0.9 keV and 3.0 keV, respectively.
In Table \ref{tab:systematics} we report, for all our systems, the 95\% upper limit on the mass cooling rate obtained by linking all cooling components below the threshold value.

\section{Systematics}\label{sec:sys}

\begin{table*}
\centering
\caption{\label{tab:systematics}Best-fit upper limits at the 95\% confidence level for $\dot M$ obtained from a modified multi {\tt mkcflow}
model (see text for details) compared to estimates of systematic effects on $\dot M$.}
\begin{tabular}{|c|c|c|c|c|c|}
\hline
Name   &   region     & E$_{\rm thr}$ & $\dot M$ best fit   & $\dot M$ free $N_{H}$  & $\dot M$ 3\% eff Area \\
       &     kpc      & keV           & $M_\odot$yr$^{-1}$  & $M_\odot$yr$^{-1}$        & $M_\odot$yr$^{-1}$     \\
\hline
A1835            & 10 &   1.8         & 17.5                &    23.3                   &      $2.1$            \\
                 & 20 &   1.8         &  9.9                &    15.6                   &      $7.1$            \\
\hline
 RXCJ1504.1      & 20 & 3.0           & 10.0                &    19.0                   &      $11.4$ \\
                         & 40 & 3.0           &  8.0                &    15.5                   &      $27.0$\\
\hline
RX J1532.9+3021  & 10 & 1.8           & 49.2                &    63.2                   &      $3.8$ \\
                 & 20 & 1.8           & 67.3                &   105.9                   &      $11.3$\\
\hline
A1068            & 20 & 0.9           & 10.0                &    14.1                   &   1.9   \\
                         & 40 & 0.9           &  4.8                &     7.4                   &   5.8\\
\hline
ZW3146           & 20 & 1.8           & 45.0                &    48.5                   &   $9.0$ \\
                         & 30 & 1.8           & 34.0                &    58.7                   &  $14.0$\\
\hline
Z0348            & 10 & 1.8           & 21.3                &   28.2                    &   $1.1$\\
                 & 20 & 1.8           & 47.9                &   74.8                    &   $4.8$\\
\hline
\end{tabular}
\end{table*}

Even in presence of a significant mass cooling rate of several tens $M_\odot$ yr$^{-1}$,
the emission associated with the cold gas contributes about 10-20\% of the total emission
in the soft band.  Clearly, our measurements are critically dependent on any effect that may
alter the signal in this energy range.
In this section we  discuss several aspects that may affect our results, showing in detail how our estimates or
upper limits on the mass cooling rates change with relevant parameters.

\subsection{Statistical uncertainties in the Galactic absorption}

The first parameter that can affect the soft emission is the Galactic absorption $N_{H}$.  In our
fitting method we keep $N_{H}$ frozen to the value found in \citet{2005LAB} at the position of the cluster.
To investigate the effect of a different value of $N_{H}$ on our results, we  repeat some of our fits with a varying
Galactic absorption.  This is motivated by the possible presence of unnoticed fluctuations in
the Galactic neutral hydrogen column densities on scales smaller than the resolution of
\citet{2005LAB}.  The typical fluctuations expected on  scales of $\sim $ 1 arcmin are on  the order of
10-40\% \citep{2003Barnes}.  Clearly, the absorption is too weak to be measured independently during
the fit.  Generally, the effect of leaving the Galactic absorption free to vary is that of  pushing $N_{H}$ towards
high values to accommodate more soft emission.
Therefore, the best-fit values or the upper limits on $\dot M$
will generally increase whenever $N_{H}$ is a free parameter.  In order to achieve a very conservative
upper limit to the amount of cold gas, we assume a maximum positive variation of 40\% in $N_{H}$.

In Table \ref{tab:systematics} we list the $\dot M$ values obtained in each cluster with a multi {\tt mkcflow} model
where all components up to a threshold value of E$_{\rm thr}$, also reported in the table, are linked.
For comparison we also report values of $\dot M$ obtained by performing the same fit with $N_{H}$ fixed to the
value found in \citet{2005LAB}.

For A1835 the largest admissible value for the Galactic
absorption is $N_{H} =  2.86 \times 10^{20}$ cm$^{-2}$.  The new upper limits are somewhat higher
reaching  $\dot M \sim 23 M_\odot$ yr$^{-1}$ and $\dot M \sim 16 M_\odot$ yr$^{-1}$ respectively
for the 10 kpc and 20 kpc regions.  Even in this case, the
upper limits on the mass cooling rate are almost an order of magnitude lower than the SFR in the BCG.
For RXCJ1504, the maximum value of  $N_{H} = 8.3 \times 10^{20}$ cm$^{-2}$ provides mass cooling rates
twice as large as in the case with $N_{H}$ frozen to its nominal values. As for A1835, this upper limit is
about an order of magnitude lower than the SFR in the BCG.
For RXJ1532, allowing $N_{H}$ to vary leads to a mass cooling rate of about 100$M_\odot$ yr$^{-1}$,
which is comparable to the SFR.
For A1068, $N_{H}$ is allowed to range up to  $2.35 \times 10^{20}$ cm$^{-2}$, and in this case the mass
cooling rates increase by less than a factor of 2, leaving the upper limit  to
less than a tenth of the SFR.
For ZW3146, the values of $\dot M$ are roughly twice as large as when $N_{H}$ is fixed, and the discrepancy with the
SFR is relatively modest.
Finally, in the case of Z0348, $N_{H}$ is allowed
to range up to  $3.5 \times 10^{20}$ cm$^{-2}$, bringing the new $\dot M$ upper limits in agreement with the SFR.


Therefore, while the value of $N_{H}$ clearly affects the measurement of $\dot M$, our results
change only moderately if we allow $N_{H}$ to increase up to a maximum of 40\% of its value measured by \citet{2005LAB}.
In addition, it is clearly very unlikely that all the measurements of $N_{H}$ for our clusters are systematically
underestimated by this amount.  We conclude that our results are not significantly affected by uncertainties
in $N_{H}$.

\subsection{Systematic uncertainties in the effective areas}

A robust characterization of weak and cool  components in a hot environment requires
that systematic effects on the effective areas be properly taken into account.  Indeed,
these components emerge only in a restricted  low-energy range, below
1 keV, where their contribution to the total flux may be quite
small. Clearly, when the relative intensity of the component becomes comparable to the
systematic uncertainties in the effective areas, any detection based  on our
maximum-likelihood analysis of the data is not reliable.
Since a fully self-consistent treatment of systematic uncertainties in the analysis of X-ray
spectra has yet to be devised \citep[see however the relevant work by][]{2006Drake},
we  resort to a rather crude but
effective ad hoc technique.  For each spectrum we take as a systematic uncertainty
in the reconstruction of the spectral shape in the 0.5-0.9 keV band,
a value of 3\% of the total count rate in that band. Thus, the systematic error on
$\dot M$ is defined as the value that the normalization of the {\tt mkcflow} model
assumes when its 0.5-0.9 keV count rate equals 3\% of the total count rate.
We apply this only to the single {\tt mkcflow} model since it would be unrealistic to
apply the entire 3\% uncertainty to a limited temperature range, and leave the other components unaffected.

We report the $\dot M$ associated with a 3\% uncertainty in the normalization of the {\tt mkcflow} model
in Table \ref{tab:systematics}.
Clearly the effect rises rapidly with the size of the extraction region, since
it is proportional to the absolute emission in the 0.5-0.9 energy range.
This is found to be lower than our upper limits in most cases,
the exceptions are the 20 and 40 kpc regions of RXJ1504 and the 40 kpc region of A1068.
We also note  that if we include the systematics on  $N_{H}$ in our comparison, the only
region to be dominated by the effective area systematics is the 40 kpc region of  RXJ1504.
In this case the difference between a cooling rate upper limit and SFR is reduced to a factor of
roughly 5.
We conclude that systematic effects on the effective area  1) do not
affect the main conclusion of this work and 2) are in most instances subdominant, implying that
further accumulation of data will lead to stronger constraints on the mass cooling rate.

\subsection{Best estimates and comparison with SFR}

\begin{figure*}
\begin{center}
\includegraphics[width=9cm]{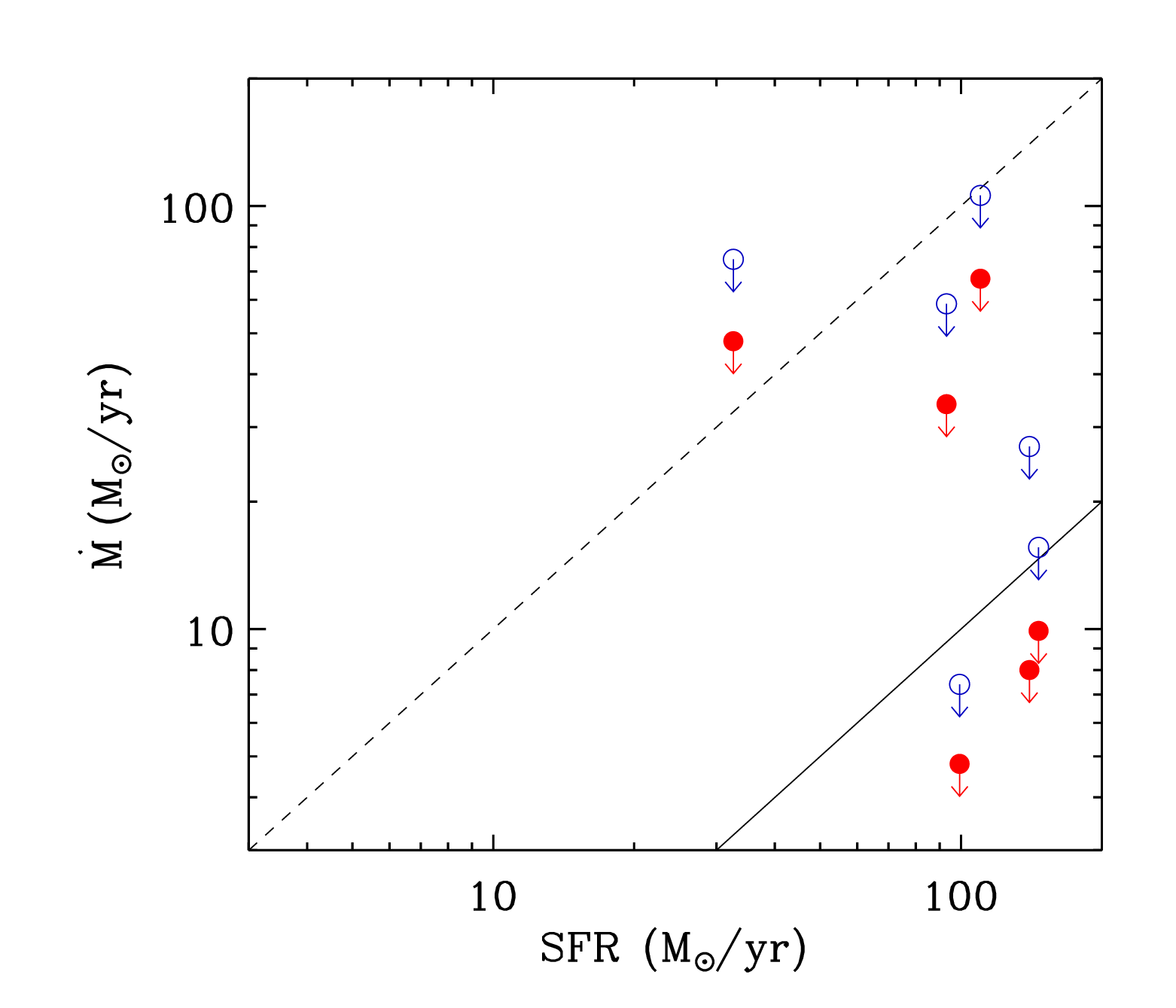}
\caption{\label{sfr_mdot_nh}Mass cooling rate vs star formation rate. Filled red circles refer to the statistical 95\% confidence upper limits on mass cooling rates. Empty blue circles refer to upper limits on cooling rates that include systematic errors.  Dashed and solid lines show the relations $\dot M =$ SFR and
$\dot M = 0.1$ SFR, respectively.}
\end{center}
\end{figure*}

Having evaluated what we consider to be the most relevant systematics,  we are now in a
position to provide a final estimate on mass cooling rates for our systems.
We do this by considering  for each system the cooling rate coming from the larger of
the two regions we have considered, these have been chosen to include all known star formation regions and,
in general, to provide more conservative upper limits.
 A statistics only estimate is provided by the values reported in the
fourth column of Table  \ref{tab:systematics}. An estimate including systematic effects is
obtained by taking the largest between the values reported in  columns  5 and 6.
We note that in only one instance, namely RXCJ1504.1, is the dominant systematic associated with
the effective area. Finally, a comparison of both estimates  (i.e., excluding  and including
systematic effects) with SFR measures is provided in Fig \ref{sfr_mdot_nh}.
As shown, in some cases, the upper limits are comparable with the star formation rate; however, in
two instances, the mass cooling rate is well below the SFR. Indeed, in the case of A1835 and A1068,
the upper limits, even allowing for systematic uncertainties, are a factor of ten lower than the
SFR. If we consider only statistical uncertainties the number of such systems goes up to 3
 (i.e., half of our sample). It is also worth bearing in mind that we do not have a cooling
rate detection for any of the systems in our
 sample, all well-known extreme star forming objects.

\section{Discussion} \label{sec:dis}

Over the past decade, a widely accepted picture has
emerged where the cooling of the bulk of the gas in cores is prevented by some form of
heating related to feedback  from the
massive black hole residing at the center of these systems.
The general idea is that small amounts of gas do manage to cool down;
one part of this gas ends up forming stars, while  the other part accretes
onto the black hole, thereby producing feedback effects. In this scenario when, occasionally, feedback
effects completely shut down  accretion onto the black hole, the hot gas  experiences
a rapid cooling, a sort of short-lived cooling flow phase, which is immediately
followed by vigorous star formation and refueling of the black hole,
thereby restarting the cycle. We note that the cooling flow phase must be short-lived, otherwise
the spectroscopic signature of the cooling gas would have been detected in some of the
many systems where it has been searched.


In this framework, the systems we have selected, i.e., those experiencing  the most
 significant star formation events, should also be those undergoing
the most intense cooling in soft X-rays.
Regrettably, this is not what we find: none of our systems show evidence of gas cooling below
a threshold temperature. Moreover, in several instances  -- four or five out of six, depending on whether we include systematic effects in our estimates --  our upper limits are significantly below the
star formation rates. In two or three cases -- depending on whether we include systematic
effects in our estimates --  we have upper limits on cooling rates that are an order of magnitude below the SFR.

 Two kinds of solutions to this problem can be considered.  In the first, the
gas is provided by some source which is not the ICM. \citet{2011Voit}, for instance,
proposed  stellar mass-loss from the old stellar population; however, in this case the
expected rate is only on the order of $10 M_\odot$ yr$^{-1}$ \citep[see also criticism in][]{2012Rawle},
which is significantly smaller than the SFR in all the systems we are considering here.

In the second class of solutions it is postulated that the
gas is supplied from the hot phase.
In this context, one way to explain the mismatch, is to assume there is a delay between the cooling
and the star formation. If this is  the case, the systems
we have chosen may have already concluded their most significant cooling phase. We now investigate
this possibility in some detail.
The hot X-ray gas must first  cool down to $\sim 10^4$ K, which is the first thermally stable region.
This takes $\sim t_{cool}$, which  is several 100 Myr for the X-ray gas. Second, the warm gas must
condense to 10-20 K, i.e., the molecular  cloud  regime.  This  is  relatively fast  ($< $ 1  Myr)
if  there  are  some  significant perturbations. Third, at the molecular stage,
the gas will start forming stars at a given rate that depends on the dynamical timescale,
$t_{dyn} = (3\pi /32 G \rho)^{1/2}$, which, in our case, should be on the order of 10-50 Myr.
It is true that $t_{dyn}$ is quite short; however, in  many astrophysical environments,
the efficiency with which stars are formed  is observed to be quite small, i.e., only a few percent
of what might be expected if $t_{dyn}$ were the relevant timescale \citep[e.g.,][and references therein]{2015Federrath}.
If these small efficiencies  apply to the core regions of BCGs, there might be a latency time of
roughly 1 Gyr and objects with very high star formation rates, such as the ones considered
in this paper, may not be the ones currently undergoing the most intense cooling in X-rays.
To address this issue, we estimate directly from existing data the depletion timescale, $t_{dep}$,
i.e., the timescale over which the molecular gas is turned into stars. We define
$t_{dep} \equiv M_{\rm mol}/{\rm SFR}$, where $M_{\rm mol}$ is the mass of molecular gas and SFR
is the star formation rate. We have measures of the molecular gas for a few of our systems
(A1835, RXJ1532, A1068, and ZW3146; see Table 1 for details) and for one system that has purposely been
left out of our study: the Phoenix Cluster \citep{2015bTozzi}. For the four systems in our study we find depletion timescales of  roughly 350 Myr for A1835 and A1068
and 1 Gyr for ZW3146 and RXJ1532, showing that the latency time for at least two of these systems is
 relatively short. Interestingly, for the Phoenix Cluster, the depletion timescale is only 30 Myr, i.e., about a factor of 10 smaller than for A1835 and A1068. The reason for much of this  difference lies in the different CO to H$_2$ conversion factor adopted by different authors. As pointed out in \citet{2014McDonald}, all these systems harbor similar amounts of molecular gas.  If the CO measurements  of  A1835 and A1068 are converted into H$_2$ mass estimates using the same conversion factor adopted for the Phoenix Cluster, the depletion timescales drops below 100 Myr, leaving essentially no latency time between X-ray cooling and star formation.
While a detailed discussion on what conversion factor should be adopted for our systems is
beyond the scope of this paper, we do wish to note that according to several recent works
\citep[see][for a review]{2013Bolatto}, the conversion factor anti-correlates with
the surface density of gas and stars. In general in BCGs the surface density is high; for instance, in the case of A1835,
in the innermost kpcs, it is on the order of 2000$M_\odot$pc$^{-2}$ \citep{2014McNamara}. This suggest that,
for our systems, a low conversion factor, such as the one adopted for the Phoenix Cluster, may indeed be appropriate.
It therefore seems that the two systems for which the upper limit on the cooling rate is a factor of 10
smaller than the star formation, namely A1068 and A1835, are also those for which the delay time
argument does not provide an adequate explanation for this discrepancy.
Of course, although it is unlikely, we might be observing A1835 and A1068 in the short time interval between the shutting off of the cooling and the consequent depletion of the molecular gas.

Another possibility is that the gas that is forming stars in our systems originates in
a significantly larger volume than the one where it currently resides.
If we assume that within a region where the cooling time is relatively short, e.g.,  several Gyr,
corresponding to scales of several tens of kpc, and some of the gas does indeed manage to cool down to a cold phase, it will increase its density significantly as it contracts and, in the absence of countering  forces,  fall towards the center of the cluster under its own weight.
Over the last few years, several high-resolution 3D simulations have been carried out to study in
detail  the  multi-phase  condensation  mechanism  we  have  sketched  here
\citep[e.g.,][]{2012Gaspari,2013Gaspari,2015bGaspari} which is known as “chaotic cold accretion” (CCA).
Such a precipitation mechanism has been  corroborated  by  independent  observational  evidence
\citep[e.g.,][and  references therein]{2015Voit}. Within this scenario, our X-ray cooling rates
might be reconciled with star formation rates if we increase the size of the regions over which
we search for the  cooling gas.
We have tested this possibility on the two systems where the upper limits on the cooling rates
are most constraining, namely A1835 and A1068 (see Table \ref{tab:systematics}). We extended our
spectral analysis for both to 100 kpc, where the cooling time  is roughly 5 Gyrs.
In neither case did we find evidence of emission of gas cooling below 1.8 keV. We estimated
upper limits of roughly  $60  M_\odot $ yr$^{-1}$ for A1835 and $40  M_\odot $ yr$^{-1}$ for A1068;
in both cases a dominant contribution to the upper limit comes from systematic errors (see \S\ref{sec:sys}),
more specifically the error associated with the effective area for A1835 and a combination of effective area and galactic absorption for A1068. These limits are still a factor of 2 below the star formation rates. Thus,
extending our search out to large radii alleviates the tension between cooling and star formation rates, but does not dissolve it entirely.
A further possibility, along these lines, is that gas with very short
cooling time, $<$ 1Gyr, exist out to significantly larger radii or perhaps throughout the cluster.
Indeed, given the limited efficiency of mechanism such as thermal conduction and mixing, a moderate degree of multi-phaseness or ``clumping'' as it is sometimes called, cannot be ruled out in the ICM at large \citep[e.g.,][and  references therein]{2016Molendi}.
We also note  that emission from the infalling cold gas may well be below the detection threshold of available instrumentation.



Finally, we  should not dismiss the possibility that, for some unknown reason, the cooling
through the soft X-ray band is either hidden from us or simply does not occur. Solutions
along these lines were proposed more than a decade ago \citep{2001Fabian}. More recently
\citep{2012Fabian} has pointed out that cold gas could be  rapidly growing by mixing with the hot gas.





\section{Summary}

We computed the isobaric mass cooling rate for a small sample of X-ray clusters observed with
{\sl Chandra}, whose BCGs are observed to host among the most significant episodes of star formation
with SFR in the range $30-150 M_{\odot}$ yr$^{-1}$.
To this end, at variance with several previous studies, we considered primarily regions
close to where  star formation is  known to occur.
Our results are summarized as follows.

 All our measurements are upper limits: in  no instance do we detect gas cooling all through the soft X-ray band. In two cases, the upper limit on the cooling rates are comparable to the SFR; in two more they are significantly below.
In the last two cases, namely  A1835 and A1068, even allowing for systematic uncertainties, they are a factor of ten lower than the SFR.

Possible systematic effects related to poor knowledge of the Galactic column density or to uncertainties
in the {\sl Chandra} calibration may affect our measurement of $\dot M$  by about a factor of  2. However,
in most of our systems, their magnitude is too modest to reconcile the discrepancy between the upper limit on the mass cooling rate and the SFR.

 We have examined several possible solutions to the cooling vs star formation rate
discrepancy.
\begin{itemize}
\item  An origin of the gas other than the ICM appears implausible as there is no obvious source.
\item In two of our systems a delay between cooling and star formation could explain the mismatch between
rates; however, for the two systems showing an order of magnitude discrepancy between star formation and cooling rates, namely A1835 and A1068, this seems unlikely.
\item A solution can be found for A1835 and A1068, if we accept that gas is
cooling out of the X-ray phase in regions that are much larger than those
on which we observe the starburst. These regions need to extend beyond the
cores of our clusters possibly encompassing significant fractions of these systems.
\end{itemize}

Finally we would like to point out that
since systematics affecting many of our measurements are not severe, a wider investigation on a
larger sample of clusters, possibly with deeper data, could provide further insight into the
relation between cooling flows and star formation events in the BCGs.

\begin{acknowledgements}
We acknowledge financial contribution from contract PRIN INAF 2012 (``A unique dataset
to address the most compelling open questions about X-ray galaxy clusters”).
M.G. is supported by NASA through Einstein Postdoctoral Fellowship Award Number PF-160137
issued by the Chandra X-ray Observatory Center, which is operated by the SAO for and on behalf
of NASA under contract NAS8-03060.

\end{acknowledgements}

\bibliography{references_Clusters_PT}

\end{document}